\documentclass[a4paper,fleqn,usenatbib,useAMS]{mnras}

\usepackage{graphicx}	
\usepackage{amsmath}	
\usepackage{amssymb}	
\usepackage{multicol}        
\usepackage{bm}		
\usepackage{pdflscape}	

\usepackage[T1]{fontenc}
\usepackage{ae,aecompl}

\usepackage{mathptmx}

\title[SDSS IV MaNGA - metallicity and nitrogen abundance gradients] 
{SDSS IV MaNGA - Metallicity and nitrogen abundance gradients in local galaxies}

\author[F. Belfiore et al.] 
{Francesco Belfiore$^{1,2}\thanks{Email: fb338@cam.ac.uk}$,
Roberto Maiolino$^{1,2}$, 
Christy Tremonti$^{3}$, 
Sebastian F. S\'anchez$^{4}$,
\newauthor
Kevin Bundy$^{5}$,
Matthew Bershady$^{3}$,
Kyle Westfall$^{5}$,
Lihwai Lin$^{6}$,
Niv Drory$^{7}$,
\newauthor 
M\'ed\'eric Boquien$^{8}$,
Daniel Thomas$^{9}$
and Jonathan Brinkmann$^{10}$
\\
\\$^1$ University of Cambridge, Cavendish Astrophysics, Cambridge, CB3 0HE, UK.
\\$^2$ University of Cambridge, Kavli Institute for Cosmology, Cambridge, CB3 0HE, UK.
\\$^3$ University of Wisconsin - Madison, Department of Astronomy, 475 N. Charter Street, Madison, WI 53706-1582, USA.
\\$^4$ Instituto de Astronom\'\i a, Universidad Nacional Auton\'oma de M\'exico, A.P. 70-264, 04510 M\'exico, D.F., Mexico.
\\$^5$ University of California Observatories - Lick Observatory, University of California - Santa Cruz, 1156 High St. Santa Cruz, CA 95064, USA.
\\$^6$ Institute of Astronomy and Astrophysics, Academia Sinica, Taipei 106, Taiwan.
\\$^{7}$ McDonald Observatory, The University of Texas at Austin, 2515 Speedway Stop C1402, Austin, TX 78712, USA.
\\$^{8}$ Universidad de Antofagasta, Unidad de Astronom\'\i a, Avenida Angamos 601, Antofagasta, 1270300, Chile.
\\$^{9}$ Institute of Cosmology and Gravitation, University of Portsmouth, Dennis Sciama Building, Portsmouth, PO1 3FX, UK.
\\$^{10}$ Apache Point Observatory, P.O. Box 59, Sunspot, NM 88349, USA.}

\pubyear{2017}

\begin{document}

\date{Accepted . Received ; in original form }
\pagerange{\pageref{firstpage}--\pageref{lastpage}} \pubyear{2016}

\maketitle
\label{firstpage}

\begin{abstract}
We study the gas phase metallicity (O/H) and nitrogen abundance gradients traced by star forming regions in a representative sample of 550 nearby galaxies in the stellar mass range $\rm 10^9-10^{11.5} M_\odot$ with resolved spectroscopic data from the SDSS-IV MaNGA survey. 
Using strong-line ratio diagnostics (R23 and O3N2 for metallicity and N2O2 for N/O) and referencing to the effective (half-light) radius ($\rm R_e$), we find that the metallicity gradient steepens with stellar mass, lying roughly flat among galaxies with $\rm log(M_\star/M_\odot) = 9.0$ but exhibiting slopes as steep as -0.14 dex $\rm R_e^{-1}$ at $\rm log(M_\star/M_\odot) = 10.5$ (using R23, but equivalent results are obtained using O3N2). At higher masses, these slopes remain typical in the outer regions of our sample ($\rm R > 1.5 ~R_e$), but a flattening is observed in the central regions ($\rm R < 1~ R_e$). In the outer regions ($\rm R > 2.0 ~R_e$) we detect a mild flattening of the metallicity gradient in stacked profiles, although with low significance. The N/O ratio gradient provides complementary constraints on the average chemical enrichment history. Unlike the oxygen abundance, the average N/O profiles do not flatten out in the central regions of massive galaxies. The metallicity and N/O profiles both depart significantly from an exponential form, suggesting a disconnect between chemical enrichment and stellar mass surface density on local scales.  
In the context of inside-out growth of discs, our findings suggest that central regions of massive galaxies today have evolved to an equilibrium metallicity, while the nitrogen abundance continues to increase as a consequence of delayed secondary nucleosynthetic production. 

\end{abstract}



\section{Introduction} 
\label{intro}

The current theoretical framework of galaxy evolution requires galaxies to be continuously fed by cosmological gas accretion onto their host dark matter haloes \citep{Fraternali2008, Almeida2014}. Halo gas subsequently cools and contributes to the growth of the disc \citep{White1978, Mo1998}, where cold gas may undergo further gravitational collapse, leading to star formation \citep{Kennicutt1998}.
Energetic feedback from stars and active galactic nuclei (AGN) is needed in order to self-regulate star formation on
galactic scales and to reproduce the overall properties of the observed galaxy population at z$\sim 0$
\citep{Croton2006, Bower2006, Vogelsberger2014, Schaye2015}. 

Decades of observational work have now demonstrated the
ubiquitous presence of galactic winds, affecting all phases of the interstellar medium (ISM, \citealt{Heckman2000,
Martin2005, Veilleux2005, Tremonti2007, Feruglio2010, Sturm2011, Cano-Diaz2012, Maiolino2012, Cicone2014}), both in the
local Universe and at high redshift. The relation between the different outflowing gas phases, however, remains
ill-characterised, together with the eventual fate of the gas - which may be either lost to the intergalactic medium
(IGM) or remain bound in the galaxy halo. The self-regulation of star
formation in galaxies therefore appears to be a complex problem, critically dependant on a tightly coupled network of exchanges of matter and energy between the different phases of the galaxy's interstellar medium (ISM) and the hot halo gas, which are challenging to observe and simulate \citep[e.g.][]{Oppenheimer2010}.

Metals are direct products of stellar nucleosynthesis, thus making chemical evolution studies a powerful complementary
approach to understanding star formation and gas flows in and out of galaxies. 
The radial distribution of metals in galaxies is of particular interest because of its direct relation to the unknown radial profile of gas accretion, which in term drives the inside-out growth of discs \citep{Larson1976, Matteucci1989, Boissier1999, Pezzulli2016}. Models of the metallicity gradient, which typically assume a simple radial dependence of the accretion rate (leading to inside-out growth) and no radial flows are capable of reproducing the observed metallicity gradients \citep{Chiappini2001, Fu2009}. 
Individual models, however, vary substantially in their most fundamental predictions. Some models, for example, predict a steepening of the metallicity gradients with time \citep{Chiappini2001, Mott2013}, while others predict the opposite \citep{Molla1997, Prantzos2000, Fu2009, Pilkington2012a}. Observations of high-redshift galaxies may provide more direct evidence regarding the time evolution of the metallicity gradients \citep{Jones2013, Wuyts2016}. Current studies, however, are still limited to small sample sizes.

Alternatively, abundance ratios of elements of different nucleosynthetic origin can be used to gain further insight into
the history of star formation and chemical enrichment in galaxies. The ratio between oxygen (mostly
produced on short timescales by massive stars dying as core-collapse supernovae) and iron (mostly produced on long
timescales by Type Ia supernovae) or nitrogen (produced by low and intermediate mass stars, \citealt{Renzini1981,
Vincenzo2016}) can be used as probes for the average timescale of star formation in a stellar system. 

Observationally it is well-known that in the local Universe disc galaxies present a negative metallicity gradient \citep{Vila-Costas1992, Oey1993, Zaritsky1994, vanZee1998, Moustakas2010}. Only recently, however, sufficiently large samples of galaxies have become available to securely quantify the slope of the metallicity gradient \citep{Moran2012, Sanchez2014, Pilyugin2015, Ho2015, Menguiano2016} and study its dependence on other galactic properties. Earlier studies, based on long-slit spectroscopy, suffered from poor spatial sampling and difficulty in excluding regions which are not dominated by ionisation from massive stars. Integral field spectroscopy (IFS) of star forming galaxies, as demonstrated by the a new generation of IFS galaxy surveys \citep{Rosales-Ortega2010, Sanchez2014, Menguiano2016, Perez-Montero2016}, has been shown to be an ideal tool to circumvent these difficulties and provide accurate measurements of chemical abundances over the whole optical extent of nearby galaxies. Recent results from the CALIFA survey \citep{Sanchez2012a} have demonstrated the existence of a characteristic negative oxygen abundance gradients in nearby galaxies \citep{Sanchez2014} and have started to explore the dependence of the metallicity gradient on stellar mass and other secondary parameters \citep{Menguiano2016, Perez-Montero2016}.

In this work we present a new study focussing on the shape of the gas phase oxygen and nitrogen abundance gradient in the nearby Universe, based on IFS data from the Sloan digital sky survey IV (SDSS IV, Blanton et al., in prep.) Mapping nearby galaxies at Apache Point Observatory (MaNGA, \citealt{Bundy2015}) survey. Our aim of this work is to provide a statistically robust reference for modellers aiming to reproduce the evolution of abundance gradients by providing an accurate anchor point for the low-redshift Universe. Unlike previous work in this area, the MaNGA galaxy sample is approximately flat in stellar mass and its simple selection function makes it possible to perform accurate volume corrections, making this the first study of the metallicity gradient in nearby galaxies to use a \textit{representative} sample of nearby galaxies over the entire mass range $\rm 10^9-10^{11.5}~M_\odot$. Moreover, this work represents a significant increase in sample size with respect to previous studies, as we derive metallicity gradients for a total of 550 nearby galaxies.

This work is structured as follows. In Sec. \ref{sec2} we introduce the MaNGA sample, data and the adopted spectral fitting procedure. In Sec. \ref{sec3} we discuss our approach for determining the gas phase oxygen and nitrogen abundances and the effect of resolution (`beam smearing') and inclination on the measured abundance gradients. In Sec. \ref{sec4} we investigate the shape of the oxygen and nitrogen abundance gradients and their dependence on other galactic properties, while in Sec. \ref{sec4A} we discuss possible sources of systematic error affecting our inference. In Sec. \ref{sec5} we discuss our findings in the context of previous studies of the shape of the metallicity gradient, while in Sec. \ref{sec6} we discuss the results in light of the theoretical picture of disc formation and chemical evolution. Finally we conclude in Sec. \ref{sec7}. 

Throughout this work we use of a $\Lambda$CDM cosmology with $\Omega_M=0.3$, $\Omega_\Lambda=0.7$ and $\rm H_0=70  ~km^{-1}~ s^{-1}~Mpc^{-1}$.



\section{The data} 
\label{sec2}

\subsection{The MaNGA data}
\label{sec2.1}

MaNGA \citep{Bundy2015} is the largest IFS survey of galaxies to date, aiming to observe a representative sample of 10~000 nearby galaxies in the redshift range 0.01 $<$ z $<$ 0.15 by 2020. The core of the MaNGA instrument \citep{Drory2015} consists of a set of 17 hexagonal fibre bundle integral field units (IFUs) of various sizes (ranging from a 19-fibre IFU, 12$''$ on sky diameter, to a 127-fibre IFU, 32$''$  on sky diameter), simultaneously deployed across the 3$^{\circ}$ diameter field of view of the SDSS 2.5m telescope at Apache Point Observatory \citep{Gunn2006}. The MaNGA instrument suite also includes a set of twelve 7-fibre minibundles used for flux calibration \citep{Yan2016} and 92 single fibres used for sky subtraction. All fibres are fed into the dual beam BOSS spectrographs covering the wavelength range from 3600 \AA\ to 10300 \AA\ with a spectral resolution R $\sim$ 2000 \citep{Smee2013}. 

The MaNGA galaxy target sample is drawn from an extended version of the NASA-Sloan catalogue (NSA {\tt v1\_0\_1}\footnote{http://www.sdss.org/dr13/manga/manga-target-selection/nsa/}, \citealt{Blanton2011}), and is selected in order to fulfil a number of design criteria motivated by the core science goals of the survey \citep{Yan2016a}. MaNGA covers galaxies out to a fixed distance in terms of effective radius ($\rm R_e$), with 2/3 of the targets selected to be covered out to 1.5 $\rm R_e$ (the `primary+' sample) and 1/3 of the targets to 2.5 $\rm R_e$ (the `secondary' sample). The MaNGA parent sample is uniformly distributed in i-band absolute magnitude, as a proxy for stellar mass (Wake at al., \textit{in prep.}). At each i-band magnitude the MaNGA primary and secondary samples are selected to be volume limited within a prescribed redshift range. 

MaNGA observations are carried out during dark time and a set of galaxies is observed until a threshold S/N value is reached \citep{Yan2016a}. A three-point dithering pattern is used during observations to compensate for light loss and create a uniform point spread function (PSF, \citealt{Law2015}). The MaNGA data used in this work was reduced using version {\tt v1\_5\_1} of the MaNGA reduction pipeline \citep{Law2016a}. The wavelength calibrated, sky subtracted and flux calibrated MaNGA fibre spectra (error vectors and mask vectors) and their respective astrometric solutions are combined to produce final datacubes with pixel size set to 0.5$''$. The median PSF of the MaNGA datacubes is estimated to have a median full width at half maximum (FHWM) of 2.5$''$. 

Stellar masses and other photometric properties for the MaNGA sample are taken from the extended NSA targeting catalogue, unless otherwise stated. The NSA stellar masses are derived using a \cite{Chabrier2003} initial mass function and using the {\tt kcorrect} software package (version {\tt v4\_2}, \citealt{Blanton2007}) with \cite{Bruzual2003} simple stellar population models.

\subsection{Spectral fitting}
\label{sec2.2}

Physical parameters of the continuum and the emission lines are obtained via a customised spectral fitting procedure described in \citealt{Belfiore2016a} (Sec. 2.2). In short, we fit the continuum using a set of simple stellar population models \citep{Vazdekis2012} and the emission lines with a set of gaussians, one per line. To increase the ability to fit weaker lines the velocities of all lines are tied together, thus effectively using the stronger lines to constrain the kinematics of the weaker ones. In the case of the [OIII]$\lambda\lambda$4959,5007 and [NII]$\lambda\lambda$6548,83 doublets their dispersions are tied together and their amplitude ratios are fixed to the ratios of the relative Einstein coefficients. 
Examples of the data products obtained with this procedure have been presented in \cite{Belfiore2016a}.

The reddening of the nebular lines is calculated from the Balmer decrement, using the $\mathrm{H \alpha  / H \beta}$ ratio and a \cite{Calzetti2001} attenuation curve with $\rm R_V = 4.05$. The theoretical value for the Balmer line ratio is taken from \cite{Osterbrock2006}, assuming case B recombination ($ \mathrm{H \alpha  / H \beta=2.87 }$). We note that the use of extinction curve of \cite{Cardelli1989} (or the modification by \citealt{O'Donnell1994}) with $\rm R_V=3.1$ yields very similar results for the 3600 $\rm \AA$ to 7000 $\rm \AA$ wavelength range considered in this work. In order to obtain a reliable extinction correction we select only spaxels with S/N $>$ 3 on both H$\alpha$ and H$\beta$.

\begin{figure} 
\includegraphics[width=0.49\textwidth, trim=0 0 0 0, clip]{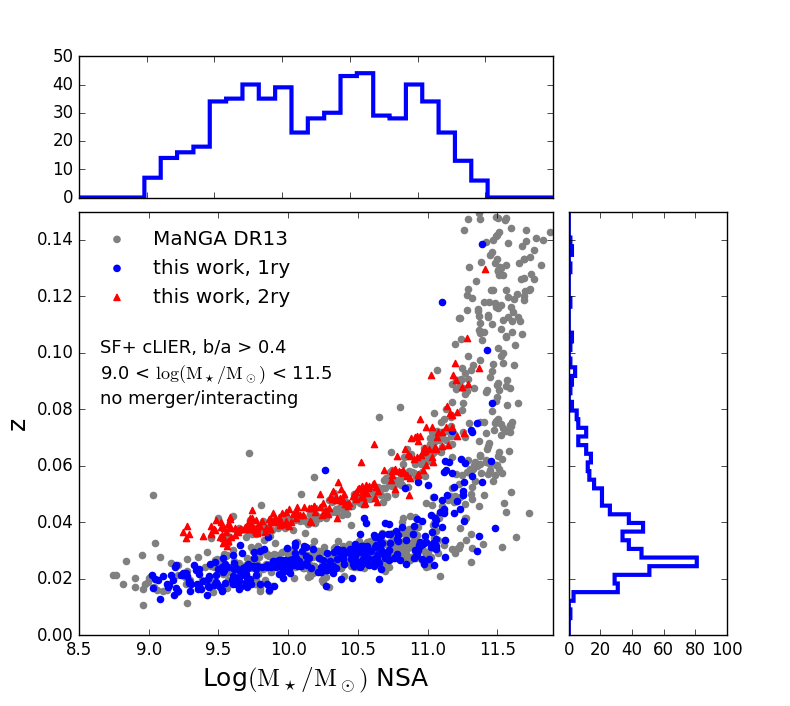}
\caption{The MaNGA sample used in this work in the redshift-$\rm M_\star$ plane. The full MaNGA DR13 sample is plotted in grey, while galaxies selected for the study of gas-phase metallicity (SF and cLIER galaxies, b/a $>$ 0.4 and $\rm 9.0 < \log(M_\star/M_\odot) <11.5$) are plotted in blue circles (primary+ sample), red triangles (secondary sample). The redshift and stellar mass distributions for the full sample used in this work are shown as blue histograms.}
\label{fig2.1}
\end{figure}

\subsection{The galaxy sample}
\label{sec2.3}

We select galaxies from a parent sample of 1392 galaxies observed by MaNGA within the first $\sim$ 2 years of operation, corresponding to the publicly available SDSS data release 13 (DR13, \citealt{SDSS_DR13}).
Following the revised galaxy classification scheme proposed in \cite{Belfiore2016a} we classify galaxies according to their emission line properties based on the [SII]/H$\alpha$ versus [OIII]/H$\beta$ Baldwin-Philips-Terlevich (BPT, \citealt{Baldwin1981, Veilleux1987, Kewley2001}) diagram as:

\begin{enumerate}
\item{\textit{Line-less galaxies}: No detected line emission (equivalent width [EW] of (H$\alpha)<$ 1.0 \AA\ within 1.0 $\rm R_e$).}
\item{\textit{Extended LIER galaxies (eLIER)}: galaxies dominated by low ionisation emission-line regions (LIERs) at all radii where emission lines are detected. No evidence for star forming regions.}
\item{\textit{Central LIER galaxies (cLIER)}: galaxies where LIER emission is resolved but located in the central regions, while ionisation from star formation dominates at larger galactocentric distances.}
\item{\textit{Star forming galaxies}: galaxies dominated by star formation in the central regions and at all radii within the galaxy disc.}
\end{enumerate}

For the remainder of this work we select galaxies with the following properties:
\begin{enumerate}
\item{Classified as star forming or cLIER based on the excitation classification from \cite{Belfiore2016a}, i.e. galaxies that display star formation at some, or all, radii.}
\item{Major to minor axis ratio (b/a) greater than 0.4, to exclude highly inclined systems.}
\item{$\rm 9.0 < \log(M_\star/M_\odot) < 11.5$ primary+ and secondary sample MaNGA galaxies.}
\item{Interacting and merging galaxies are excluded since both observations and simulations confirm that interacting and merging galaxies have systematically flatter metallicity gradients due to gas mixing \citep{Kewley2010, Rich2012}. In particular, we exclude galaxy pairs, defined as galaxies within the SDSS spectroscopic catalogue with projected separation $\rm < 70~kpc$ and line-of-sight velocity difference of $\rm < 500~km~s^{-1}$. In order to exclude late-stage mergers, where the small separation between the two galaxies prevents SDSS from having spectroscopic coverage of both galaxies, we make use of the visually classified merger catalogue from the Galaxy Zoo merger project \citep{Holincheck2016}. We exclude galaxies with a merger probability $\rm p_{merger}> 0.4$.}
\end{enumerate}

Fig. \ref{fig2.1} displays the position of the selected MaNGA galaxies in the redshift-mass plane, with the full DR13 sample also displayed in grey. As discussed in Sec. \ref{sec2.2}, MaNGA observes larger (hence generally more massive galaxies) at higher redshifts, in order for these galaxies to be covered to the appropriate galactocentric distance with the available IFUs. For the same reason, the secondary sample targets are always observed at higher redshift with respect to primary+ sample targets of the same luminosity. 

After applying the cuts described above to the MaNGA DR13 sample we obtain a sample of 550 galaxies (334 primary+ sample, 216 secondary, in blue circles and red triangles in Fig. \ref{fig2.1} respectively) for which we derive gas phase metallicity gradients. 


\section{Measuring gas phase metallicity} 
\label{sec3}

\subsection{Identifying star forming regions}
\label{sec3.1}

The formation of new stars in galaxies can be traced by observations of H\textsc{ii} regions, where photoionisation equilibrium is maintained by radiation from embedded hot massive O and B stars. Apart from classical, discrete H\textsc{ii} regions, galaxies are observed to contain low surface brightness diffuse ionised gas (DIG), sometimes also referred to as warm ionised gas \citep{Reynolds1984, Hoopes2003, Oey2007, Zhang2017}. The ionisation and energetics of DIG are a subject of active research. A combination of radiation leaking from classical H\textsc{ii} regions, massive stars in the field and radiation from hot evolved stars may all be required to reproduce the observed properties of the DIG \citep{Oey1997, Hoopes2003, Zhang2017}. In particular, spectroscopic studies of the DIG in the Milky Way \citep{Reynolds1995, Madsen2006} and external galaxies \citep{Rand1990, Rossa2003, Kaplan2016, Zhang2017} demonstrate that DIG shows emission line ratios typical of LIERs, thus requiring a combination of lower ionisation parameter and a harder ionisation field than H\textsc{ii} regions. Since metallicity diagnostics are calibrated on H\textsc{ii} regions models, it is necessary to minimise the contribution of DIG when measuring chemical abundances from emission line ratios.

H\textsc{ii} regions have typical sizes of tens to a few hundreds parsecs and are thus not resolved by MaNGA (at the median redshift of 0.03 the MaNGA PSF FWHM corresponds to 1.5 kpc). In this work we make use of diagnostic line ratios to separate star forming regions (which are assumed to be collections of unresolved H\textsc{ii} regions) and DIG. In particular, we exclude regions which are not classified as star forming using the [SII]/H$\alpha$ versus [OIII]/H$\beta$ BPT diagram and the \cite{Kewley2001} demarcation line. As demonstrated in \cite{Belfiore2016a}, this criterion successfully excludes extra-planar and inter-arm regions in star forming galaxies.

Previous work has also demonstrated that there exists a tight relation between diagnostic line ratios and EW(H$\alpha$) in emission, in the sense that LIERs have EW(H$\alpha$) $<$ 3 \AA\ in emission \citep{CidFernandes2011, Sanchez2015, Belfiore2016a}. After applying the BPT cut described above and also requiring S/N $>$ 3 on the metal lines required to measure chemical abundances, only 0.3 \% of spaxels considered have EW($\rm H\alpha$) < 3 \AA\ and 3 \% have EW($\rm H\alpha$) < 6 \AA. The exclusion of these spaxels from further analysis does not change any of the subsequent results presented in this paper.


\subsection{Measuring chemical abundances}
\label{sec3.2}

The gas phase metallicity (mass fraction of all heavy metals) is generally traced through the oxygen abundance, as oxygen is the most abundant heavy element by mass in the ISM. Following the convention in the field, in this paper we use the terms `gas phase \textit{oxygen} abundance' and  `metallicity' interchangeably
\footnote{Note, however, that this use of the term metallicity is in contrast to the general use of the term in stellar astrophysics,
where Fe abundance is often taken as a metallicity proxy.}.

Observationally the most reliable way to measure gas phase oxygen abundance, at least in the sub-solar regime, is by
determination of the electron temperature of the nebula (the so-called $\rm T_e$ method, \citealt{Pagel1992, Izotov2006}) via
detection of weak auroral lines (i.e. [OIII]$\lambda$4363, [NII]$\lambda$5755). These lines, however, are $\rm \sim
100-1000$ times fainter than $\rm H \beta$, and thus generally undetected at the typical depth of MaNGA and other comparable spectroscopic surveys (like the legacy SDSS spectroscopic survey). In absence of direct temperature measurements, strong line diagnostics ratios can be calibrated, either empirically (via samples of H\textsc{ii} regions with direct method abundances) or theoretically (making use of photoionisation models), to obtain an indirect estimate of the gas phase metallicity.

Despite recent advances \citep{Dopita2013, Perez-Montero2014, Blanc2015, Asari2016, Curti2017}, reliably measuring the gas metallicity from strong emission lines ratios remains a difficult problem in observational astrophysics. Different calibrations, even when based on the same diagnostics ratios, can give results differing by up to 0.6 dex \citep{Kewley2008, Lopez-Sanchez2012, Pena-Guerrero2012}. It is beyond the scope of this work to resolve the abundance scale problem. In order to illustrate the results obtained using different metallicity calibrations, in this work we make use of two independent metallicity calibrators. 

\begin{enumerate}
\item{The \cite{Pettini2004} (PP04) metallicity calibration based on the O3N2 index, defined as
\begin{equation}
\rm O3N2 \equiv \log \frac{[OIII] \lambda 5007/H\beta}{[NII] \lambda 6584/H\alpha}.
\end{equation}
This index has the advantage of varying monotonically with metallicity and of relying on line ratios which are very close in wavelength, thus minimising the uncertainty due to the extinction correction or imperfections in the flux calibration. The calibration is empirical in nature and is anchored to the abundances of 137 H\textsc{ii} regions (of which 6 are derived by detailed photoionisation models, and the others via the direct method) and suffers from a possible systematic uncertainty in the super-solar metallicity regime. 
It should also be noted that, since [NII] and [OIII] originate from ions with a large difference in ionisation potential, this diagnostic ratio is systematically affected by variations in the ionisation parameter (see for example \citealt{Blanc2015}), which are not calibrated. Moreover the calibration assumes a one-to-one relation between the nitrogen and oxygen abundance, which is not necessarily a good assumption in presence of complex star formation history or gas flows \citep{Henry2000, Belfiore2015, Vincenzo2016}.}

\item{The calibration from \cite{Maiolino2008} (M08), based on the R23 parameter,
\begin{equation}
\mathrm{R_{23} =  ([OII] \lambda\lambda 3726,28+ [OIII]\lambda\lambda 4959, 5007) / H \beta}.
\end{equation}
The M08 calibration is a hybrid one. At low metallicities (12+log(O/H)<8.4) it is empirically calibrated through
the $T_e$ method; in this metallicity range the $T_e$ is deemed reliable (not suffering significantly
from temperature fluctuation effects), while models struggle to reproduce the strong line ratios.
At high metallicities (12+log(O/H)>8.4) the calibration is linked to the models of \cite{Kewley2002}; in this
metallicity range the $T_e$-based metallicities are thought to be biased low as a consequence of temperature fluctuation
effects, while photoionization models reproduce the line ratios reasonably well.

In the high metallicity range
the M08 calibration gives similar abundances to those obtained using the calibration used by \cite{Tremonti2004} and the calibration of R23 presented by  \cite{Kobulnicky2004} (KK04), which solves recursively for metallicity and ionisation parameter.\footnote{In fact, the dependence of R23 on the ionisation parameter is relatively small on the upper branch of the R23-O/H relation \citep{Kewley2002, Blanc2015}, which is where most of the MaNGA data lies.}
This calibration is more sensitive than O3N2 to possible inaccuracies in the extinction correction, due to the large difference in wavelength between [OII]$\lambda\lambda$3726,28 and the other two lines. This calibration, however, is not sensitive to changes in the N/O ratio.}
\end{enumerate}


The derivation of the N/O abundance ratio from the [NII]$\lambda$ 6548, 6584 to [OII]$\lambda \lambda 3726,29$ line ratio (N2O2) is subject to smaller systematics than the measurement of the gas phase oxygen abundance. The advantage of using this line ratio lies in the similar ionisation potential of oxygen and nitrogen, which guarantees that $\rm N^+/O^+ = N/O$ to high accuracy \citep{Vila-Costas1993, Thurston1996}. \cite{Pagel1992} provides a convenient formula to relate $\rm N^+/O^+$ to the N2O2 calibrator, which has been verified theoretically using a five-atom calculation

\begin{equation}
\mathrm{\log(N^+/O^+) = \log(N2O2)+0.307 -0.02 \cdot \log{t_{NII}} - 
0.726 \cdot \log{t_{NII}}^{-1}  },
\end{equation}
where $\rm t_{NII}$ is the [NII] nebular electron temperature, which is assumed to be the same as the [OII] temperature \citep{Stasinska1990, Garnett1992}. For the purposes of this work, we estimate $\rm t_{NII}$ using the calibration provided by \cite{Thurston1996} which relies on R23. The N/O abundance ratios obtained with this calibration are equivalent to the ones obtained using the more recent \cite{Pilyugin2010} method, based on a similar set of strong lines ratios. All results are unchanged if we choose to use the \cite{Pilyugin2010}  calibration instead.


\begin{figure} 
\includegraphics[width=0.52\textwidth, trim=20 20 0 0, clip]{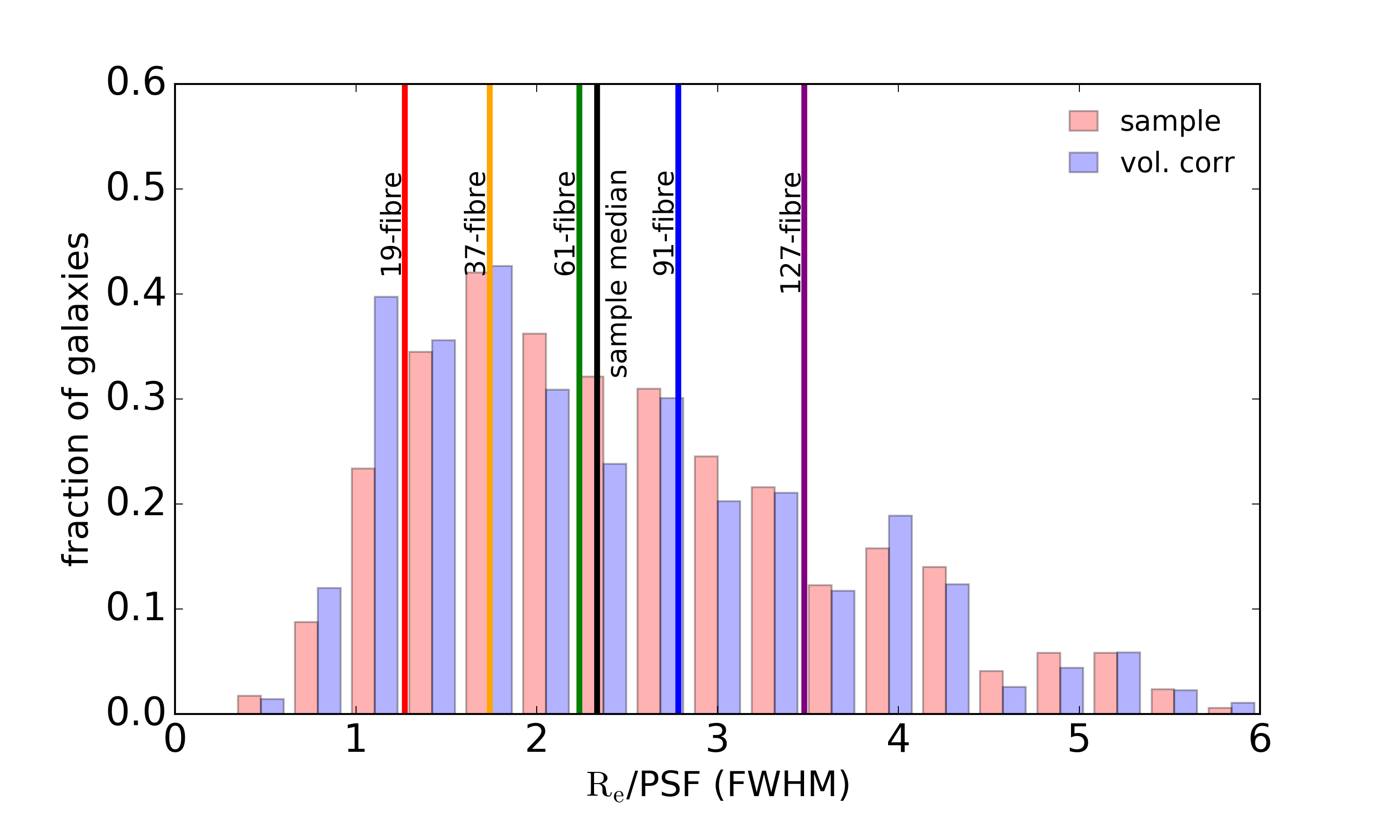}
\includegraphics[width=0.49\textwidth, trim=0 0 40 20, clip]{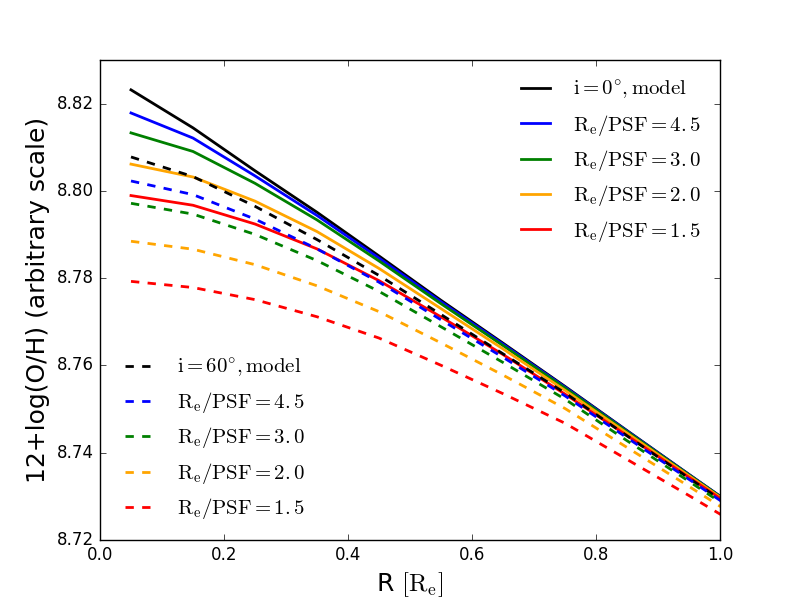}
\caption{
\textit{Top}:  Histogram of the ratio of the effective radius of MaNGA targets and the MaNGA PSF (FWHM). The red histogram represents the full MaNGA sample used in this work, while the blue histogram represents the volume-corrected distribution. Since MaNGA observes galaxies to a fixed galactocentric distance in units of $\rm R_e$ independently of the galaxy brightness, the volume correction does not introduce a large change in the underlying distribution. The coloured vertical lines represent the median values of $\rm R_e$/PSF for the whole sample ($\rm R_e$/PSF = 2.3) and for bundles of different sizes (19-fibre: 1.3 , 37-fibre: 1.7, 61-fibre: 2.2, 91-fibre: 2.8, 127-fibre: 3.5).
\textit{Bottom}: The effect of inclination and beam smearing on the model metallicity gradient, computed from a projected 3D exponential flux model for two emission lines. Solid lines refer to a face-on disc ($i=0^\circ$) while dashed lines are for $i=60^\circ$. Lines of different colour represent the effect of varying $\rm R_e/PSF$, as detailed in the legend. 
}
\label{fig3.3}
\end{figure}


\subsection{The effect of spatial resolution}
\label{sec3.3}

Observed metallicity gradients can be systematically affected by the `beam smearing' effect of the PSF. This effect is increasingly important at higher redshift and for marginally resolved galaxies \citep{Yuan2013, Mast2014, Stott2014, Wuyts2016} and also affects the measurement of the metallicity gradient at the typical resolution of the MaNGA data. In the MaNGA survey galaxies are observed out to a fixed galactocentric radius in units of $\rm R_e$, hence MaNGA bundles of different sizes have systematically different $\rm R_e/PSF$ ratios: larger bundles afford more resolution elements per $\rm R_e$. By virtue of the MaNGA target selection, to first order there is no correlation between the galaxy brightness (or stellar mass) and the size of the bundle it has been allocated. This means that, on average, $\rm R_e/PSF$ does not depend on stellar mass within the MaNGA sample (see also Sec. \ref{sec4A.2}). This conclusion does not hold if one is interested in physical distance, i.e. $\rm kpc/PSF$, which depends systematically on mass within the MaNGA sample, as more massive galaxies are observed at higher redshift.

In Fig. \ref{fig3.3}, bottom panel, we show the distribution of the $\rm R_e/PSF$ ratio for the MaNGA sample used in this work (red histogram). The median $\rm R_e/PSF$ is 2.3 (varying from 1.3 for the 19-fibre bundles to 3.5 for the 127-fibre bundles). Applying a volume correction (blue histogram) does not have a large effect on the $\rm R_e/PSF$ distribution, since $\rm R_e/PSF$ is to first order independent of galaxy brightness by selection. 

In order to study the effect of beam smearing in a model situation we construct a 3D model of an exponential disc with vertical scale height corresponding to an oblateness of q=0.13, as assumed in Sec. \ref{sec3.4}. We then rotate the model to different inclinations ($\rm i = 0, 30, 45, 60, 75^{\circ}$) and generate 2D on-sky projections by summing the flux along the line-of-sight. For each projection, we convolve the resulting 2D image with a Gaussian PSF with FWHM corresponding to $\rm R_e/PSF = 1.5, 2.0, 3.0, 4.5$, in order to cover the range of parameter space relevant to the MaNGA data. Since metallicity is a function of line ratios we perform this procedure for two different flux maps (corresponding in this example to the [NII] and [OIII] lines), chosen to generate a metallicity gradient with a slope of -0.1 $\rm dex~R_e^{-1}$. Since $\rm R_e$ is the only spatial scale in the problem the beam smearing effect depends only on the ratio between $\rm R_e$ and the resolution element ($\rm R_e/PSF$) and the assumed slope of the metallicity gradient.

Fig. \ref{fig3.3}, bottom panel, shows the effect of beam smearing on the metallicity gradient for the $i=0^\circ$ (solid lines) and $i=60^\circ$ (dashed lines) model. The colour of the lines represents different amount of beam smearing, while the black lines representing the result from using the un-convolved 2D projected models. From this simulation we conclude that

\begin{enumerate}
\item{The effect of beam-smearing is increasingly important at smaller galactocentric radii for all inclinations.}

\item{Even with no beam smearing, in this simple model, inclination generates a flattening of the metallicity gradient, since flux from different galactocentric radii is summed when the model is projected in the plane of the sky. A residual dependence of the mass-metallicity relation on inclination, likely caused by this effect, is indeed observed in SDSS single-fibre abundances \citep{Tremonti2004}. Our toy model does not take into account differential extinction, which would further attenuate the flux from the central, metal-rich regions and produce an even more significant flattening of the metallicity gradient.}
\end{enumerate}

For each of the models produced we fit a linear metallicity gradient and study the effect of inclination and beam smearing on the recovered slope for models of different inclinations as a function of $\rm R_e/PSF$. Since the central regions of galaxies are the most affected by beam smearing, we attempt a fit to the recovered metallicity gradient in two different radial ranges: (1) between 0.5 and 2.0 $\rm R_e$ and (2) over the whole considered radial range (0.0-2.0 $\rm R_e$).

As expected, we observe a systematic flattening in the metallicity gradient with decreasing $\rm R_e/PSF$ and also with increasing inclination, even for large values of $\rm R_e/PSF$. This remark motivates the exclusion of highly inclined systems (b/a <0.4, or $i > 68^\circ$) in this work. Moreover, the inclusion of the central regions ($\rm R< 0.5~R_e$) in a linear fit to the metallicity gradient causes a significant systematic flattening of the derived slope. We therefore recommend the exclusion of the central regions in order to recover the intrinsic metallicity gradient in the MaNGA sample. The systematic bias in the recovered metallicity gradient with respect to the true gradient increases for steeper model gradients (in units of $\rm dex~R_e^{-1}$), but the qualitative conclusions of this sections remain valid.

The toy model presented above does not aim to reproduce the details of the nebular emission in real galaxies. Several important factors are not taken into account, including the clumpy and non-uniform flux distribution of line emission, the effect of dust (and thus of differential extinction across the galaxy disc) and the systematic change in the property of line emission from classical H\textsc{ii} regions to the diffuse ISM \citep{Zhang2017}. The latter effect, in particular, implies that star forming regions are increasingly contaminated by lower surface brightness DIG as the physical resolution of the observations is degraded. These possible systematic effects should be taken into account when comparing the results derived in this work with observations derived at much higher spatial resolution. Further analysis of the resolution effects in the MaNGA sample is presented in Sec. \ref{sec4A.2}.

\subsection{Deriving the abundance gradient}
\label{sec3.4}

Following the classification of spaxels as star forming (Sec. \ref{sec3.1}), the gas phase metallicity is derived using the calibrations described in Sec. \ref{sec3.2} for each spaxel in which the relevant line ratios are detected with S/N $>$ 3. To determine the radial abundance gradient we calculate the deprojected distance of each spaxel. The inclination (i) is derived from the measured semi-axis ratio (b/a) by assuming a constant oblateness of q = 0.13 \citep{Giovanelli1994} according to
\begin{equation}
\rm \cos^{2}(i) = \frac{(b/a)^2-q^2}{1-q^2}.
\end{equation}
We have checked that the inclinations measured in this way are consistent those derived from a thin disc modelling of the MaNGA velocity field, as further discussed in Appendix \ref{app:B}. 

The choice of the correct scale length to normalise the abundance gradient in disc galaxies is not obvious. We seek a definition of scale length which is both robust and physically motivated. To this aim, we adopt the elliptical Petrosian effective radius (henceforth $\rm R_e$), which is the most robust measure of the photometric properties of MaNGA galaxies provided by the NSA catalogue. We discuss in Sec. \ref{sec4A.1} the effect of choosing a different normalising radial scale length.


\begin{figure*} 
\includegraphics[width=0.49\textwidth, trim=0 0 0 0, clip]{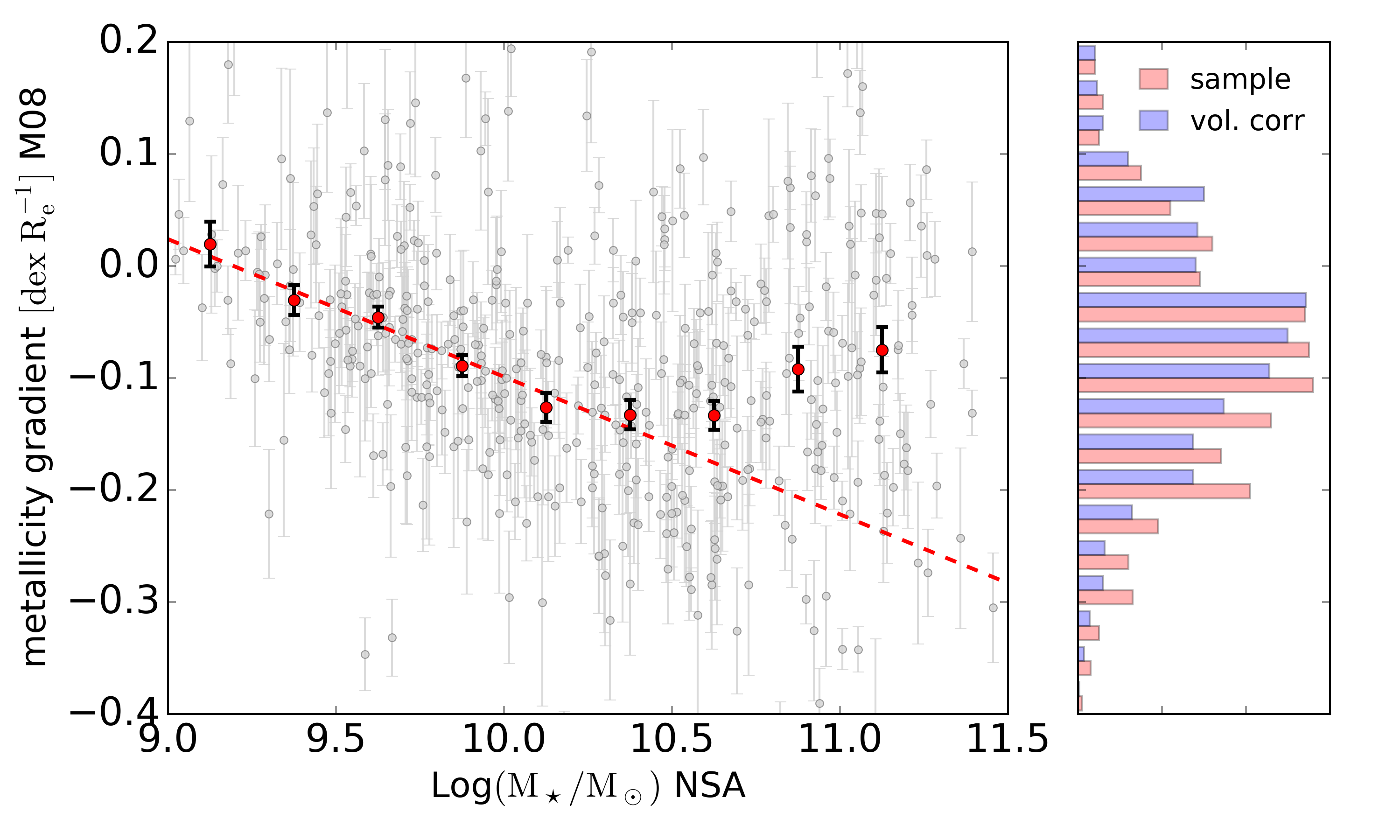}
\includegraphics[width=0.49\textwidth, trim=0 0 0 0, clip]{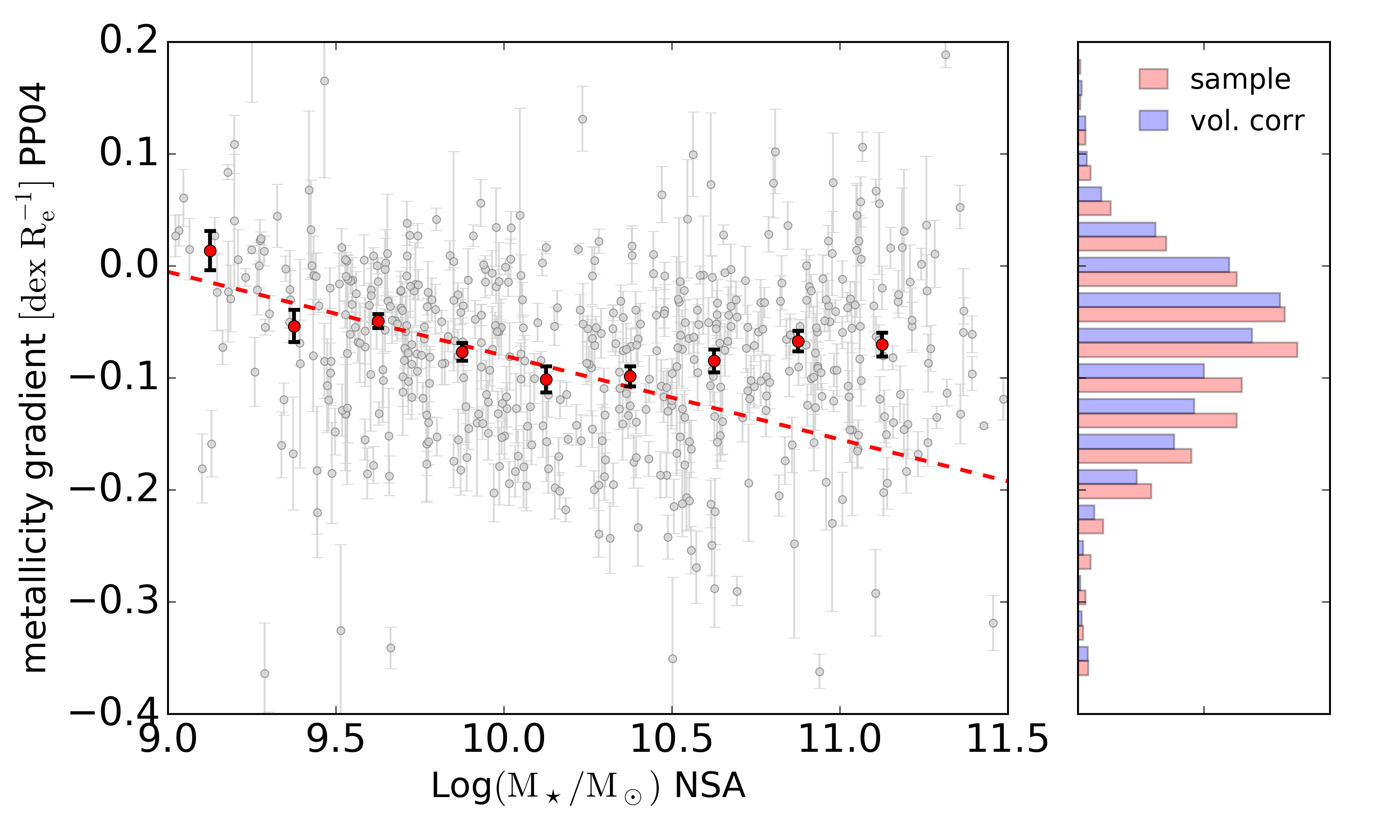}
\includegraphics[width=0.49\textwidth, trim=40 0 40 20, clip]{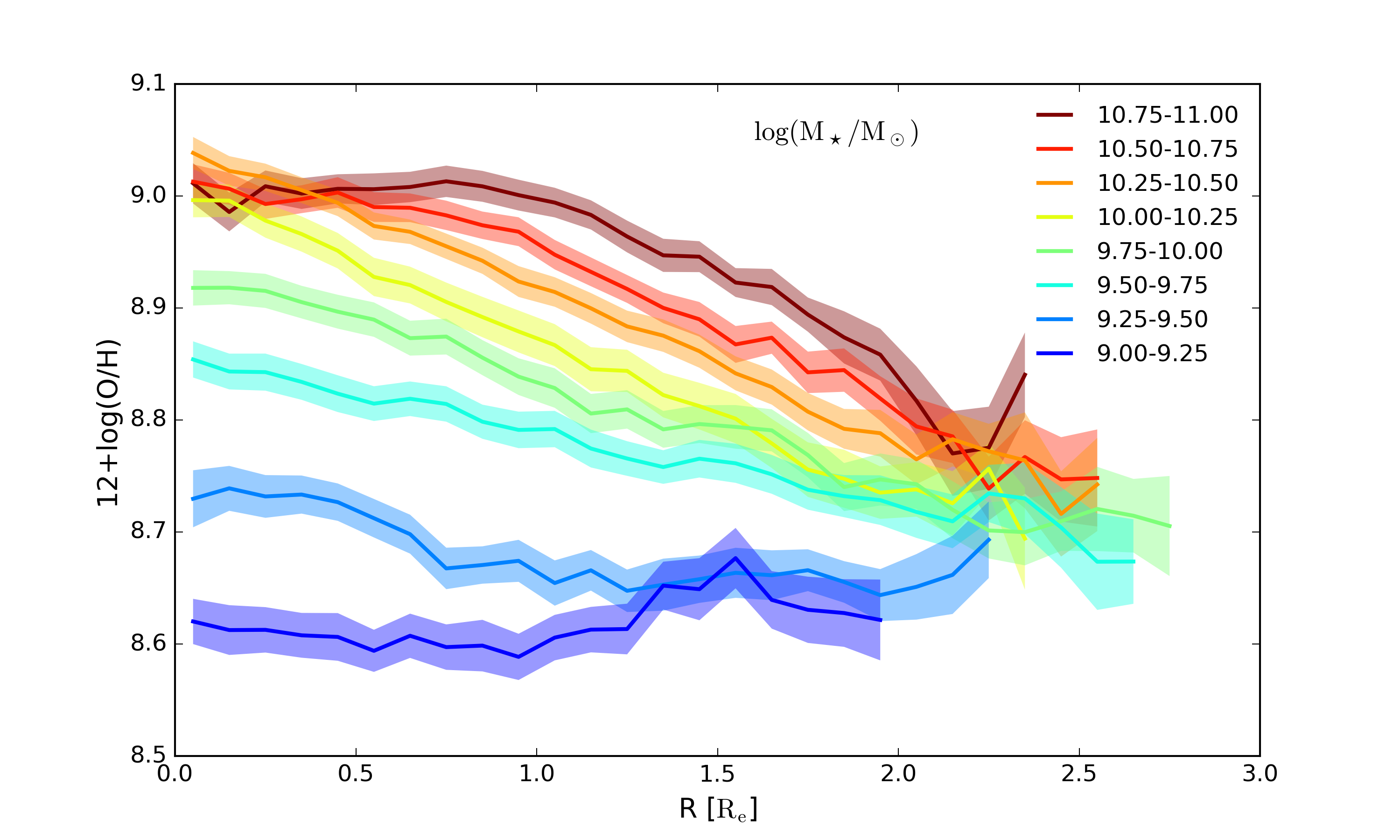}
\includegraphics[width=0.49\textwidth, trim=40 0 40 20, clip]{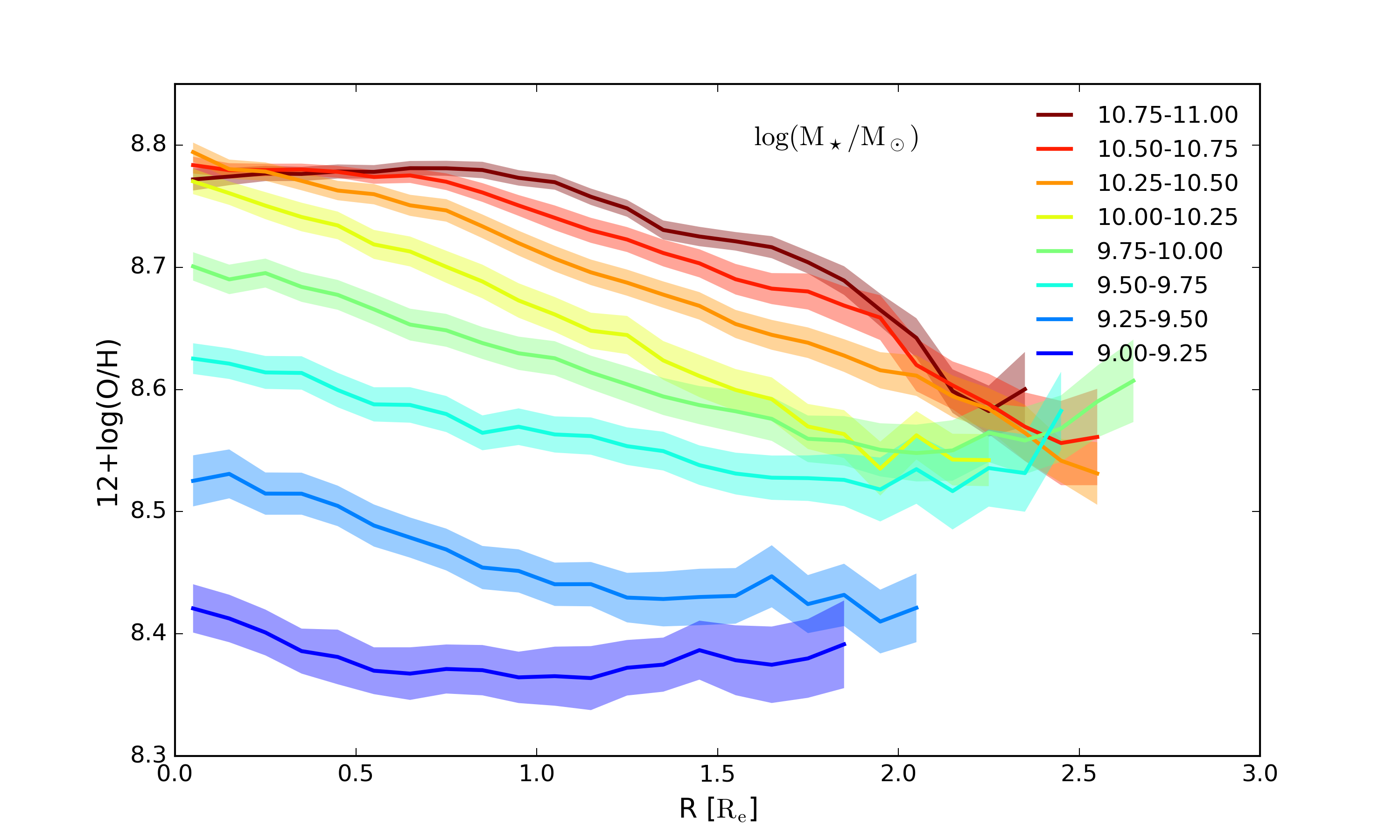}

\caption{\textit{Top}: The metallicity gradient (measured in the radial range 0.5 - 2.0 $\rm R_e$) using the M08 calibration based on R23 (left) and the PP04 calibration based on O3N2 (right) as a function of stellar mass. The red points represent the median measurements in 0.25 dex stellar mass bins with corresponding errors. The red dashed line is a straight line fit to the median gradient as a function of mass for $\rm log(M_\star/M_\odot) < 10.5$. The histogram on the right represents the distribution of metallicity gradients for the full sample (red) and for a volume weighted sample (blue). 
\textit{Bottom}: The shape of the metallicity gradient (using the M08 calibration based on R23, on the left, and PP04 calibration based on O3N2, on the right) in 0.25 dex mass bins from $\rm log(M_\star/M_\odot) = 9.0 -11.0$. For each mass bin the shaded region represents the error on the median gradient.}

\label{fig4.1}
\end{figure*}

For each galaxy we compute the benchmark metallicity gradient in the radial range 0.5-2.0 $\rm R_e$, unless otherwise stated (although not all galaxies are covered to $\rm 2.0~R_e$). This radial range is used to minimise the effects of inclination and beam smearing on the gradient measurements (see Sec. \ref{sec3.3}), but also in light of the fact that significant deviations from a linear fit are observed at both smaller and larger radii \citep{Sanchez2014,  Menguiano2016}. 

In order to fit a straight line model to the metallicity gradient, we derive the azimuthally averaged metallicity in radial bins of 0.1 $\rm R_e$ and perform an unweighted least-squares linear fit. A gradient is not computed for galaxies with less than four valid metallicity data points in this radial range 0.5-2.0 $\rm R_e$. The errors on the metallicity gradient are derived by a bootstrapping analysis. The residuals from the initial line fit are drawn randomly and with replacement and applied to data, which is then refitted with a straight line model. For each galaxy the process is repeated 1000 times and the standard deviation of the bootstrapped gradient is taken to be the formal error.

\section{The shape of the oxygen and nitrogen abundance  gradients}
\label{sec4}

\subsection{The mass dependence of the metallicity gradient}
\label{sec4.1}

Fig. \ref{fig4.1} (top left) shows the dependence of the metallicity gradient on stellar mass. The metallicity gradient is measured in the radial range 0.5 - 2.0 $\rm R_e$ with the M08 R23 calibration. Individual gradients are plotted as grey points, while the median metallicity gradient in 0.25 dex bins of stellar mass is plotted as red points with associated errors on the mean.

We observe that the slope of the oxygen abundance gradient becomes increasingly negative with stellar mass, going from roughly flat at $\rm log(M_\star / M_\odot) = 9.0 $ to $\rm -0.14~dex~R_e^{-1}$ at $\rm log(M_\star / M_\odot) =10.5$. In the stellar mass range $\rm 9.0< log(M_\star/M_\odot) < 10.5$ the relation between the slope of the gradient and $\rm \log(M_\star)$ is well fitted by a straight line with negative slope (red dashed line). Galaxies of higher mass show an inversion of this trend, displaying shallower, but still negative gradients. 

In Fig. \ref{fig4.1} (top left) we also show the histogram distribution of all the measured gradients (red) and the derived distribution after applying the volume weights (blue), thus representing our inference for the gradient distribution in a volume-limited sample. The mean and scatter of the metallicity gradient distribution for the MaNGA sample is $\mu = -0.08$ and $\sigma = 0.12$ using the M08 calibration. In the case of a volume-limited sample (in the stellar mass range $\rm 10^9-10^{11.5}~ M_\odot$), low mass galaxies, which display flatter gradients, are preferentially up-weighted, thus causing an overall flattening of the volume-averaged metallicity gradient, which we infer to have mean and scatter $\mu = -0.06$ and $\sigma = 0.12$. 
An equivalent trend between metallicity gradient and stellar mass is observed using the \cite{Pettini2004} metallicity calibration based on O3N2 (see Fig. \ref{fig4.1}, top right panel). With this calibration the mean and scatter of the metallicity gradient for the MaNGA sample is $\mu = -0.08$ and $\sigma = 0.10$ ($\mu = -0.07$ and $\sigma = 0.10$ after volume correction).

In Fig. \ref{fig4.1} (bottom) we show the stacked profiles for the metallicity gradient (using the M08 R23 and the PP04 O3N2 calibrations respectively) for galaxies of different stellar masses. The profiles are obtained by computing the robust estimate of the median profile (using Tukey's biweight, \citealt{Beers1990}) and standard deviation within each mass bin. Shaded regions represent the error on each stacked profile, which takes into account the number of galaxies contributing at each radius. For each mass bin a profile is computed only if more than 12 galaxies have a valid measured profile at that radius. 

The mass-metallicity relation is responsible for the zero-point vertical offset between the different profiles. Aside from the vertical offset, the figure highlights the differences in \textit{shape} of the metallicity profiles for galaxies of different stellar masses. In particular, we observe that
\begin{enumerate}
\item{Galaxies with  $\rm log(M_\star/M_\odot) < 9.5$ have flatter profiles than higher mass galaxies in the full radial range probed by our observations.}
\item{Galaxies with  $\rm log(M_\star/M_\odot) > 10.5$ show a flattening in the central regions ($\rm R<0.5 ~R_e$). While these are the regions of the metallicity profile most affected by beam-smearing, they are also flatter with respect to the inner regions of lower mass galaxies. Since $\rm R_e/PSF$ does not depend on mass, the observed flattening is highly significant in a differential sense.}
\end{enumerate}
The same qualitative features regarding the shape of the metallicity profile are observed by making use of the \cite{Pettini2004} metallicity calibrator, despite the differences in normalisation of the oxygen abundance scale.

The choice of radial range considered has an effect on the measured slope of the metallicity gradient, especially for higher mass galaxies, as they show strongly flattened profiles out to $\rm 1.0~R_e$. A steeper slope of the metallicity gradient is naturally obtained if a larger fraction of the central regions is excluded from the linear fit for high mass galaxies. 

\begin{figure} 
\includegraphics[width=0.49\textwidth, trim=20 50 20 50, clip]{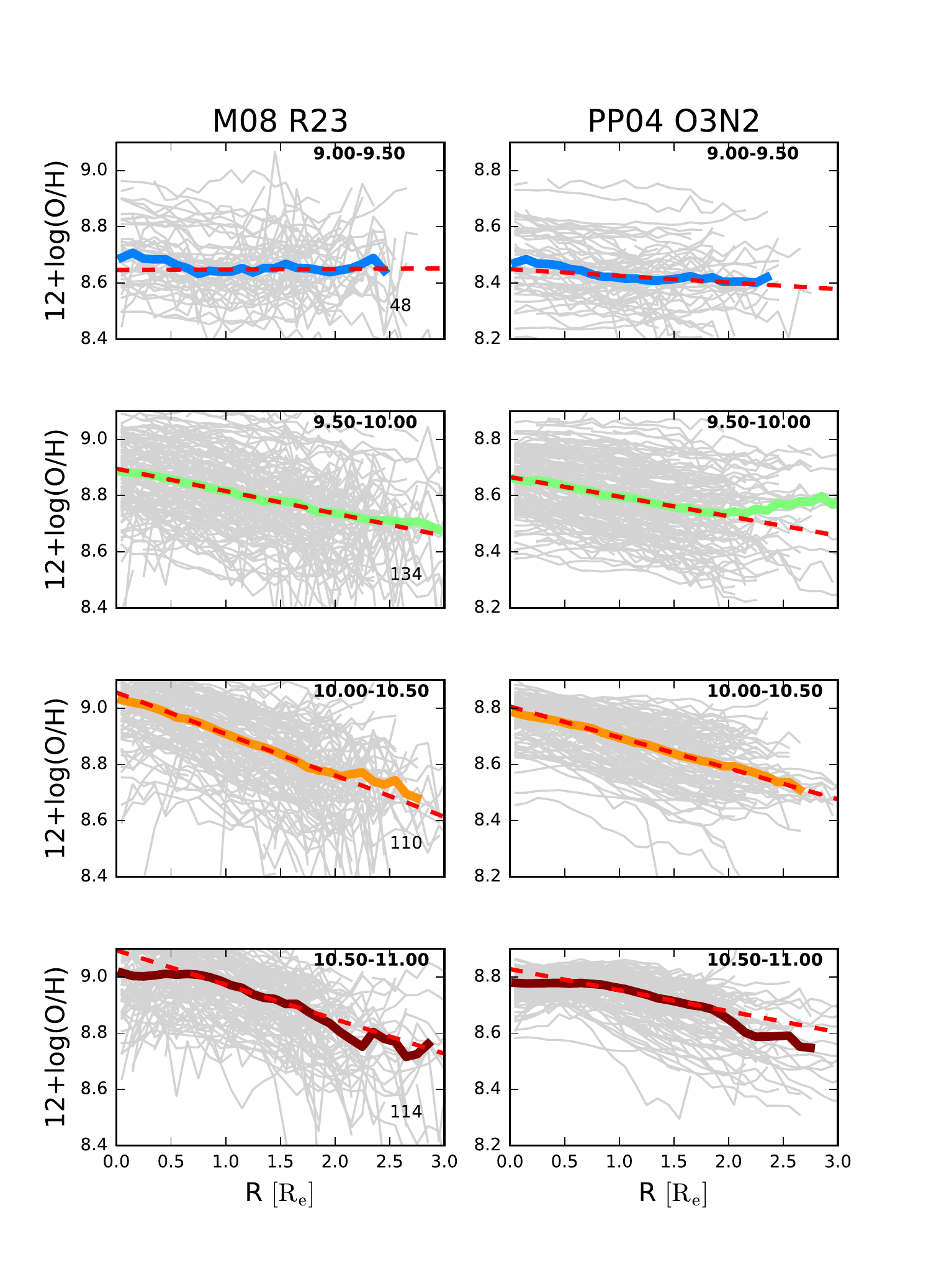}
\caption{Metallicity profiles for all the individual galaxies in the sample considered in this work (grey lines), subdivided into stellar mass bins. For each bin (mass range shown in the top right corner of each panel), the median profile is shown as a thick coloured line and the linear fit to the profile in the $\rm 0.5< R/R_e < 2.0$ radial range is shown as a dashed red line. The number of galaxies in each mass bin is shown in the bottom right corner of each panel. The analysis is presented for both the M08 and PP04 metallicity calibrators.}
\label{fig4.extra1}
\end{figure}

In order to demonstrate the distribution of observed metallicity profiles, we plot in  Fig. \ref{fig4.extra1} the metallicity gradients (grey lines) obtained for individual galaxies in four mass bins (masses shown in the top right corner of each panel). The coloured lines correspond to the median profile in the mass bin, while the red dashed line corresponds to the best fit linear model for the metallicity gradient in the radial range $\rm 0.5< R/R_e < 2.0$. The number of galaxies in each mass bin is shown in the bottom right corner, and the distributions are shown for both the M08 and PP04 metallicity calibrations. In general, the individual profiles are rather smooth, with no evidence for large breaks in metallicity.
Deviations from a linear profile in the metallicity gradient are difficult to study in individual galaxies, due to the large scatter, and are most evident in mass stacks, as discussed further in Sec. \ref{sec4.5}.

\begin{figure*} 
\includegraphics[width=0.8\textwidth, trim=50 20 80 0, clip]{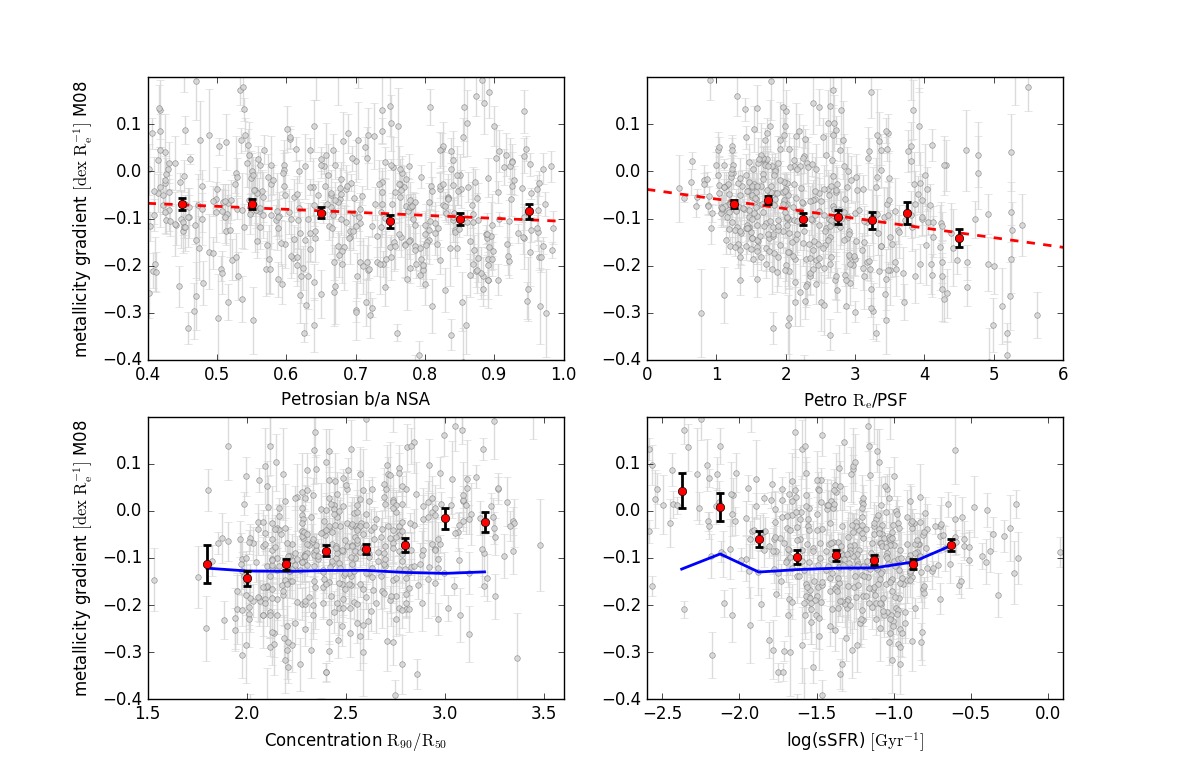}
\caption{The slope of the metallicity gradient as a function of axis ratio (b/a, from elliptical Petrosian photometry), $\rm R_e$/PSF (FWHM), as a measure of the beam smearing effect, concentration ($\rm R_{90}/R_{50}$) and specific star formation rate (sSFR $\rm [Gyr^{-1}]$) using the M08 metallicity calibrator based on R23. The red dashed lines are a linear fit to the binned data (red data points with error bars). The blue solid lines in the lower panels are the expected metallicity gradient for each bin if stellar mass was the only driving parameter, predicted using the relation between the metallicity gradient and stellar mass from Sec. \protect\ref{sec4.1}.}
\label{fig4.3}
\end{figure*}

\subsection{Dependence of the metallicity gradient on secondary parameters}
\label{sec4.3}

In Fig. \ref{fig4.3} we study the dependence of the metallicity gradient in the radial range 0.5-2.0 $\rm R_{e}$ on several secondary parameters. 

Using the current sample selection, the metallicity gradient has a weak negative dependence on inclination (parametrised by the elliptical Petrosian axis ratio, b/a) and a negative dependence on the number of PSF FWHM within one $\rm R_e$ (top panels in Fig. \ref{fig4.3}). 
By fitting a straight line relation between the slope of the metallicity gradient and b/a we obtain that the gradient steepens by 0.03 dex $\rm R_e^{-1}$ going from b/a of 0.4 to face-on galaxies. Performing the same procedure with the PSF/$\rm R_e$ dependence, we obtain that the gradient steepens by 0.08 dex $\rm R_e^{-1}$ going from 1.5 to 3.5 PSF/$\rm R_e$ (corresponding roughly to the expected resolutions for the 19-fibre and the 127-fibre bundles respectively).

In the bottom panels of Fig. \ref{fig4.3} we show the dependence of the metallicity gradient on galaxy concentration (a proxy for the bulge fraction, defined as $\rm R_{90}/R_{50}$, where R are Petrosian radii containing 90\% and 50\% of the light respectively) and specific SFR ($\rm sSFR = SFR / M_\star$). The SFR is calculated from the extinction-corrected H$\alpha$ emission for the whole galaxy, according to the method in \cite{Belfiore2017}. The data shows a positive correlation of metallicity gradient with concentration, in the sense of more highly concentrated galaxies having flatter gradients. The dependence between metallicity gradient and sSFR is flat for log(sSFR) $>$ -1.5 $\rm Gyr^{-1}$, but galaxies with lower sSFR show remarkably flatter metallicity gradients.

Both concentration and sSFR correlate with stellar mass, in the sense that galaxies with higher concentration and lower sSFR also have higher stellar mass. In order to test the relation of these secondary parameters with the metallicity gradient, it is necessary to control for the effect of stellar mass. Making use of the observed relation between stellar mass and metallicity gradient, for each bin in concentration or sSFR we calculate the slope of the metallicity gradient expected from its median stellar mass. The resulting predicted slopes are plotted as blue solid lines in the lower panels of Fig. \ref{fig4.3}. This exercise demonstrates that stellar mass is consistent with being the main driving parameter of the metallicity gradient for low concentrations and high sSFR galaxies (C $< $2.6, log(sSFR) $>$ -1.5 $\rm Gyr^{-1}$), but galaxies of high concentration and low sSFR show flatter gradients than expected from the stellar mass dependence alone. 

We have checked that the deviations from the expected gradient observed for high concentration and low sSFR galaxies persist when using the PP04 O3N2 calibration.

\subsection{The mass dependence of the N/O ratio gradient}
\label{sec4.4}

\begin{figure*} 

\includegraphics[width=0.6\textwidth, trim=0 0 0 0, clip]{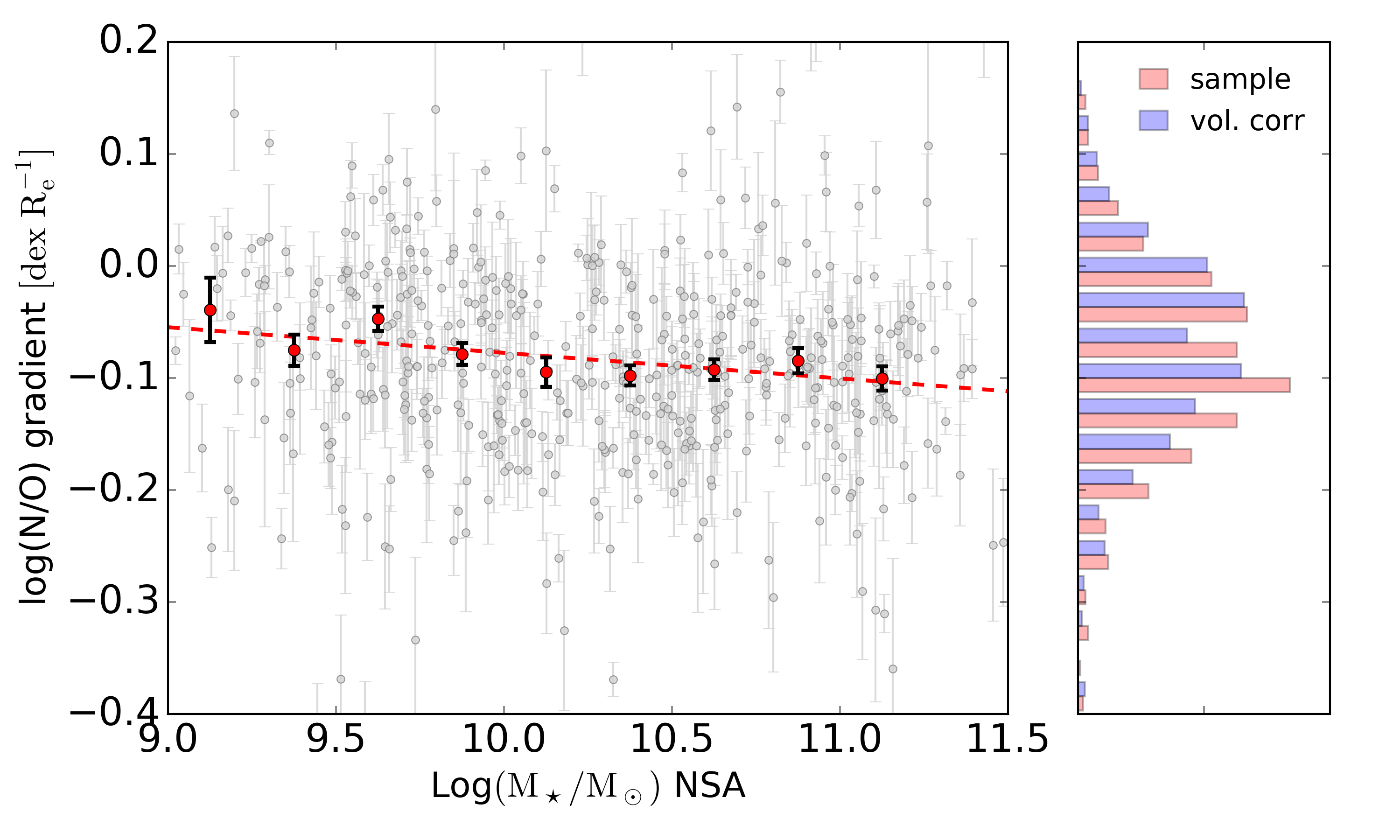}
\includegraphics[width=0.7\textwidth, trim=0 0 0 20, clip]{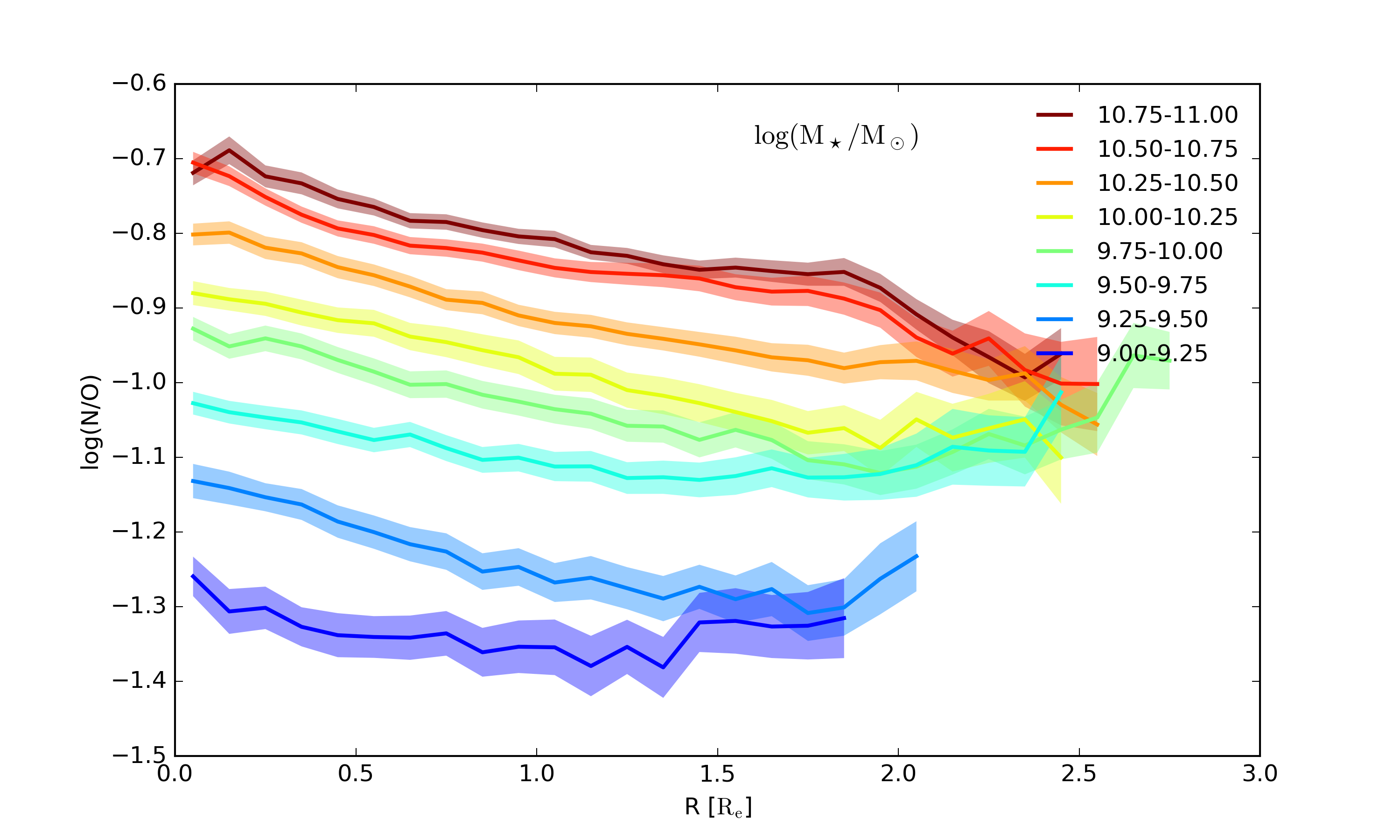}

\caption{\textit{Top}: The log(N/O) gradient (measured in the radial range 0.5 - 2.0 $\rm R_e$ using the \protect\cite{Pagel1992} calibration based on N2O2) as a function of stellar mass. The red points represent the median measurements in 0.25 dex stellar mass bins with corresponding errors. The red dashed line is a straight line fit to the median gradient as a function of mass. The histogram on the right represents the distribution of log(N/O) gradients for the full sample (red) and for a volume-weighted sample (blue). The volume-weighted distribution is fitted with a Gaussian with mean of -0.16 dex $\rm R_e^{-1}$ and $\sigma =$ 0.09 dex $\rm R_e^{-1}$.
\textit{Bottom}: The shape of the log(N/O) gradient in 0.25 dex mass bins from $\rm log(M_\star/M_\odot) = 9.0 -11.0$. For each mass bin the dashed region represents the error on the median gradient.}

\label{fig4.4}
\end{figure*}

Fig. \ref{fig4.4} (top) shows the dependence of the log(N/O) gradient (measured in the radial range 0.5 - 2.0 $\rm R_e$) on stellar mass. Individual gradients are plotted as grey points, while the median gradient in 0.25 dex bins of stellar mass is plotted in red symbols, with associated errors. The distribution of log(N/O) gradients for the whole sample (red) and for a volume-corrected sample (blue) are plotted as a histogram on the right of panel of Fig. \ref{fig4.4}, demonstrating that the log(N/O) gradient is slightly steeper than the average metallicity gradient, and is affected by similar scatter (mean log(N/O) gradient of -0.07 $\rm dex~R_e^{-1}$ with scatter of 0.09 dex). In contrast to the metallicity gradient, the log(N/O) gradient is observed to steepen with stellar mass over the whole observed stellar mass range. The steepening represents, however, only a modest effect (-0.02 $\rm dex~R_e^{-1}$ per dex in stellar mass).

Contrary to the behaviour of the metallicity gradient, the log(N/O) abundance ratio does not flatten in the innermost regions of massive galaxies (Fig. \ref{fig4.4}, bottom), where a flattening of the oxygen abundance gradient is observed. In fact, we detect a \textit{steepening} of the N/O ratio gradient in the inner regions of the most massive galaxies in the sample, as further discussed in Sec. \ref{sec4.5}.

\subsection{The behaviour at small and large radii}
\label{sec4.5}

In this section we focus on a quantitative comparison of the changes in slope of the metallicity and N/O gradients for galaxies of different masses as a function of radius. 

\begin{figure*} 
	\includegraphics[width=0.49\textwidth, trim=0 0 0 0, clip]{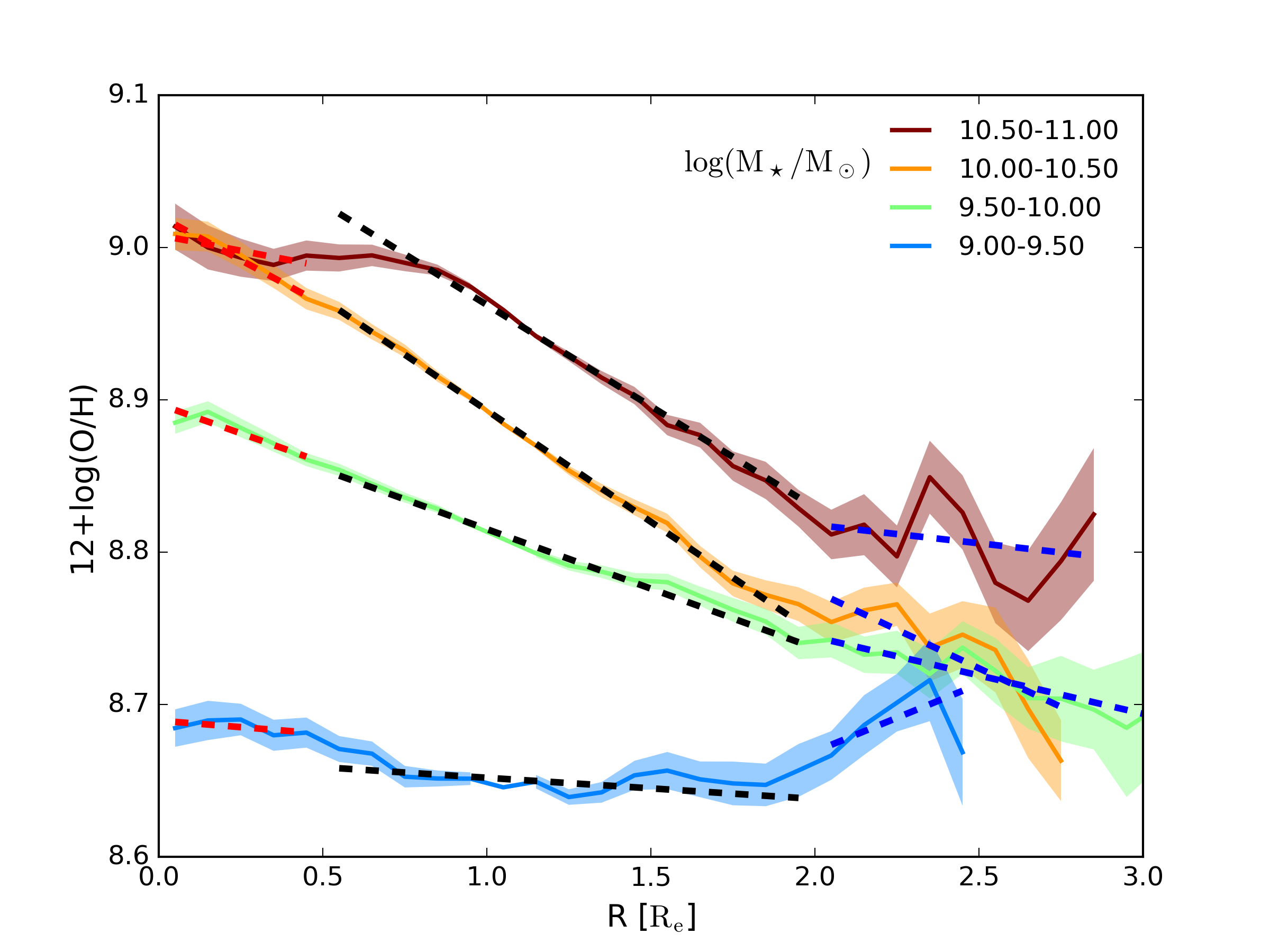}
	\includegraphics[width=0.49\textwidth, trim=0 0 0 0, clip]{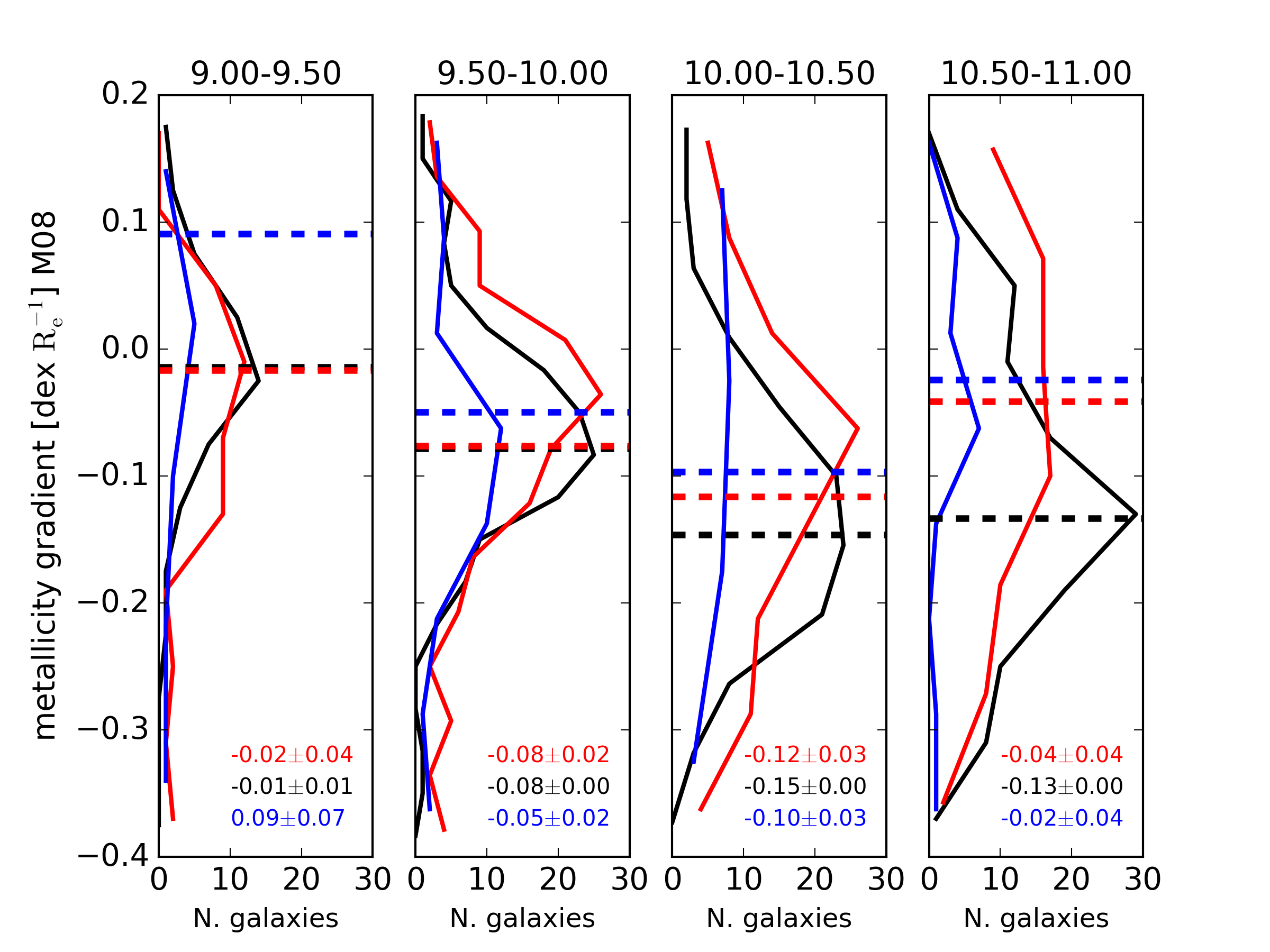}
	\includegraphics[width=0.49\textwidth, trim=0 0 0 0, clip]{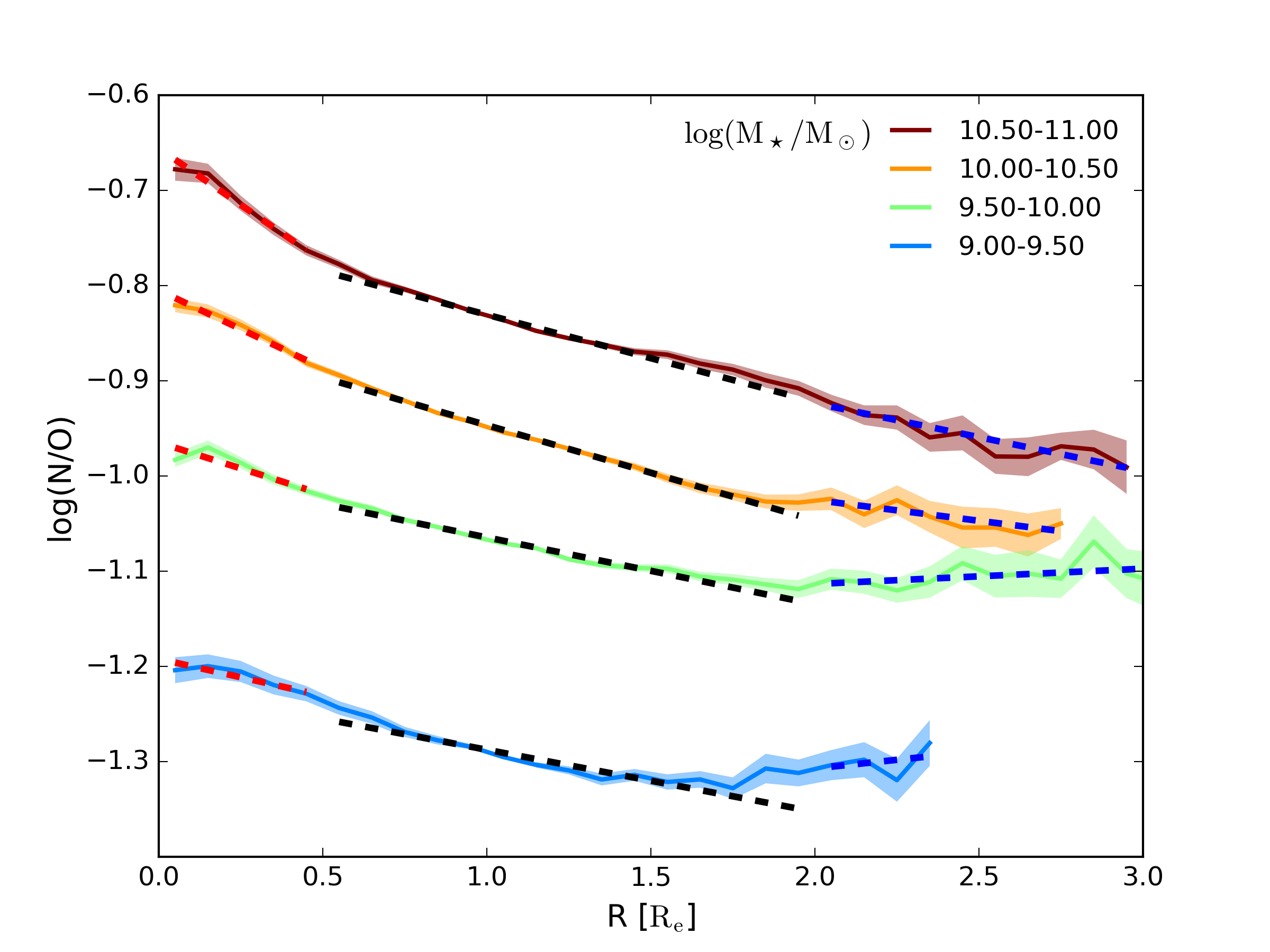}
	\includegraphics[width=0.49\textwidth, trim=0 0 0 0, clip]{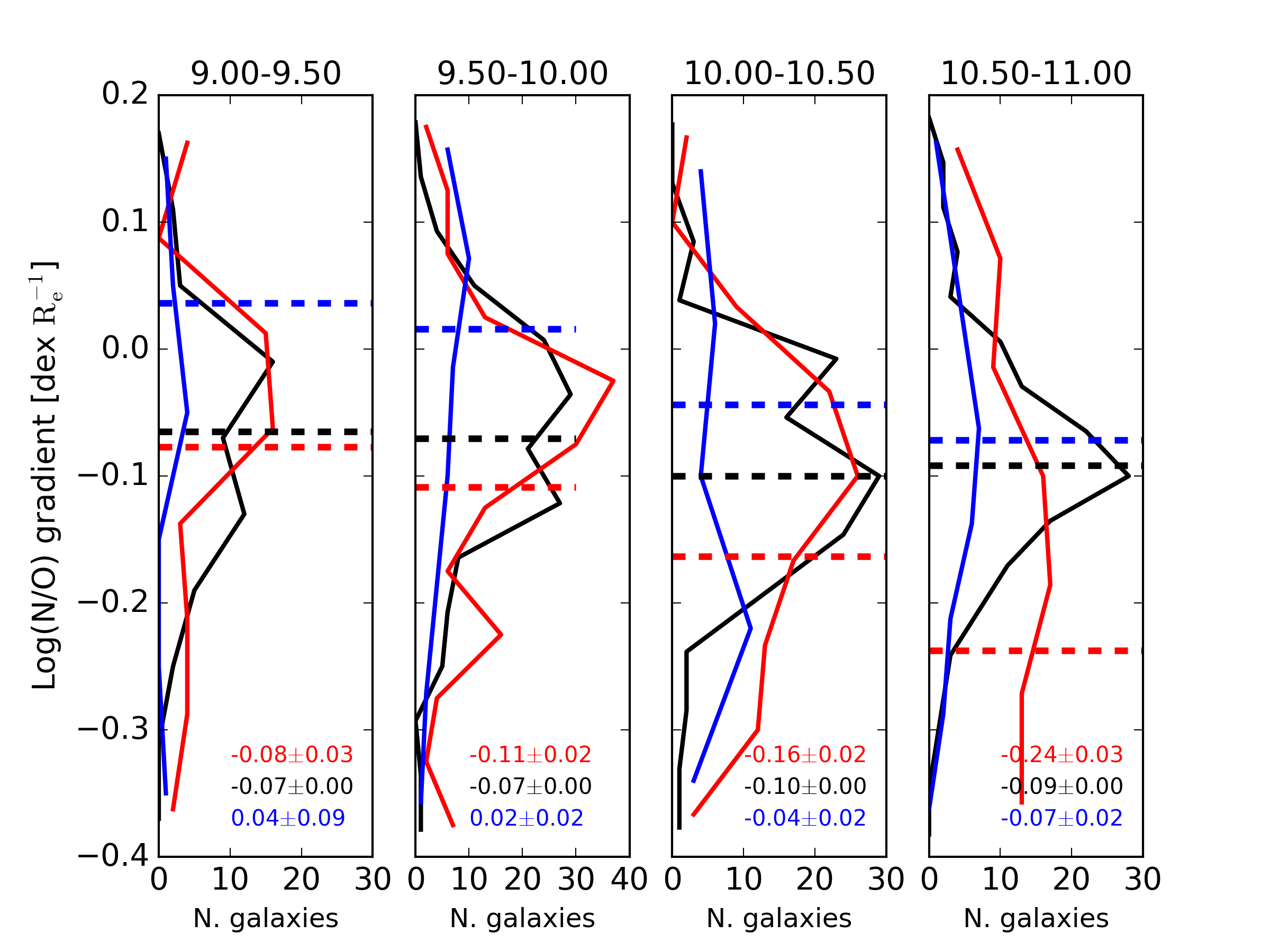}
	\caption{\textit{Top-left}: Stacked metallicity profiles in four stellar mass bins and straight line fits in three radial ranges: 0.0-0.5 $\rm R_e$ (red), 0.5-2.0 $\rm R_e$ (black) and 2.0-3.0 $\rm R_e$ (blue). \textit{Top-right}: Histogram distributions of the slope of the metallicity gradient measured in three different radial range, 0.0-0.5 $\rm R_e$ (red), 0.5-2.0 $\rm R_e$ (black) and 2.0-3.0 $\rm R_e$ (blue), for the four mass bins (mass of the bin indicated above each panel). The slopes obtained from fitting the stacks are also shown as dashed lines, and their values and error is reported in the bottom-right corner of each panel. \textit{Bottom}: Same as the top panels, but for the N/O profiles.}
	\label{fig4.extra2}
\end{figure*}

To do so, we define three radial ranges 0.0-0.5 $\rm R_e$, 0.5-2.0 $\rm R_e$ and 2.0-3.0 $\rm R_e$ respectively and study both the distribution of the metallicity profiles in individual galaxies and the results from a stacking analysis in stellar mass bins. We stack metallicity profiles in four stellar mass bins 0.5 dex wide (Fig. \ref{fig4.extra2}). Each profile is normalised to the median value of all profiles in that bin at 1.0 $\rm R_e$, in order to mitigate the effect of the mass-metallicity relation. We then fit each profile with a straight line in the three different radial ranges considered (shown as red, black and blue dashed lines in Fig. \ref{fig4.extra2}, left). The slopes obtained from fitting the stacks are compared with the distributions obtained by fitting individual galaxy profiles shown as histograms in Fig. \ref{fig4.extra2}. The red, black and blue histograms correspond to the radial ranges 0.0-0.5 $\rm R_e$, 0.5-2.0 $\rm R_e$ and 2.0-3.0 $\rm R_e$ respectively. The slopes obtained from fitting the stacks are shown as horizontal dashed lines and their value and uncertainty is quoted in the bottom right corner of each panel.

This analysis demonstrates that inner regions of massive galaxies ($\rm log(M_\star/M_\odot)$ 10.5-11.0) have flatter O/H profiles than the same regions of less massive galaxies. In particular, while for the highest mass bin the slope in the inner region is marginally consistent with being flat, for lower-mass galaxies the slope in the inner regions is statistically consistent with that measured in the 0.5-2.0 $\rm R_e$ range.

At large radii, a flattening of the metallicity gradient with respect to the value $\rm 0.5 < R/R_e < 2.0$ is detected for $\rm log(M_\star/M_\odot) > 9.5$, although with varying degrees of significance. In particular, for galaxies in the mass bins $\rm log(M_\star/M_\odot) =$ 9.5-10.0 and 10.0-10.5, the gradient at large radii is still negative and only marginally flatter than in the $\rm 0.5 < R/R_e < 2.0$ radial range. For the lowest stellar mass bin the metallicity gradient at large radii is slightly inverted, although with low significance. The observed distribution of the slopes of the metallicity gradients at large radii (as shown in the histograms of Fig. \ref{fig4.extra2} right) are broad, highlighting the advantage of using the stacked profiles. Applying a volume correction to these distributions only has a small effect in each mass bin, since the volume corrections are mass-dependent. 

The behaviour of the log(N/O) ratio gradient with mass and radius is shown on the bottom panels of Fig. \ref{fig4.extra2}. Similarly to the metallicity gradient, the log(N/O) gradient flattens at large radii, with the middle two mass bins showing the most significant flattening with respect to the slope in the $\rm 0.5 < R/R_e < 2.0$ radial range. In contrast to the metallicity gradient, the log(N/O) ratio gradient steepens in the inner regions of galaxies with increasing mass. This steepening is especially evident for $\rm log(M_\star/M_\odot) > 10.5$.

\subsection{The relation between oxygen and nitrogen abundance}
\label{sec4.6}

\begin{figure*} 
\includegraphics[width=0.7\textwidth, trim=0 0 0 0, clip]{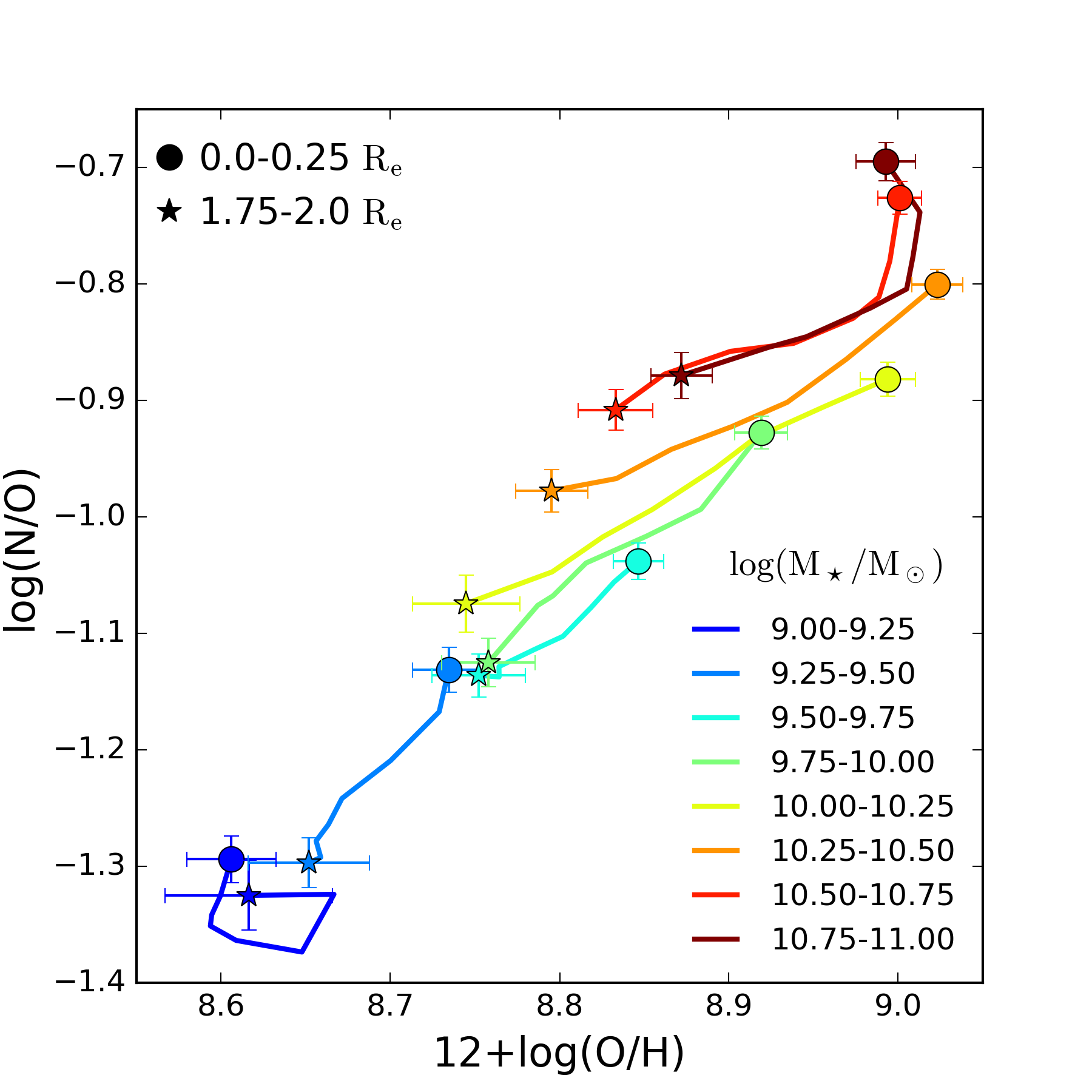}
\caption{The relation between log(N/O) and metallicity (12+log(O/H)) as a function of both stellar mass and radius. The radial profiles in both metallicity and N/O are stacked in stellar mass bins of width 0.25 dex. The coloured solid lines represent the location of the stacked radial profiles in the log(N/O) versus 12+log(O/H) plane, with circles representing the innermost radial bin (0.0-0.25 $\rm R_e$) and the stars representing the outermost radial bin (1.75-2.0 $\rm R_e$).}
\label{fig4.6}
\end{figure*}

In Fig. \ref{fig4.6} we show the relation between metallicity (using the M08 calibration) and the N/O ratio. All the
valid spaxels in the MaNGA data have been divided in 8 bins of 0.25 dex according to the total stellar mass of the galaxy. For each mass bin, spaxels have been subdivided according to galactocentric distance (normalised to Petrosian $\rm R_e$) in 8 radial bins. For each radial bin the median log(N/O) is plotted as a function of median metallicity (different colours indicate the different stellar mass bins). For each stellar mass, the innermost radial bin ($\rm 0.0-0.25~R_e$) is plotted with a filled circle and the outermost radial bin ($\rm 1.75-2.0~R_e$) with a star. Representative errors bars are plotted for these bins.

The derived relation between O/H and N/O agrees with previous work, showing a super-linear dependence of nitrogen on oxygen abundance at super-solar metallicity. A plateau is expected at log(N/O) $\sim$ -1.4 for lower metallicities, which are however not well represented in the MaNGA data. Fig. \ref{fig4.6} also confirms the results of previous sections, by demonstrating the flattening of both N/O and O/H gradients at lower stellar masses (particularly evident for the $\rm 9.0 < log(M_\star/M_\odot) < 9.25$ bin). The flattening of the metallicity gradient in the inner regions of massive galaxies, which is however {\it not} accompanied by a flattening of the N/O ratio, is also evident (for example in the $\rm 10.75 < log(M_\star/M_\odot) < 11.00$ bin, which shows an increase in N/O but not O/H moving towards the inner radii). 

The key result from Fig. \ref{fig4.6} is that the scatter in the O/H versus N/O relation correlates with galaxy mass. In particular, at a fixed local O/H, outer regions of more massive galaxies have higher N/O than inner regions of less massive galaxies, at least for $\rm log(M_\star/M_\odot) > 9.5$. Possible physical processes responsible for this effect are discussed in Sec. \ref{sec6.3}.

\section{Sources of systematic error}
\label{sec4A}

In this section we discuss possible sources of systematic error in the determination of chemical abundances and their radial gradients in the MaNGA data.

\subsection{The choice of radial scale length}
\label{sec4A.1}

Throughout this work we have used the Petrosian effective radius as the normalising scale length to measure metallicity gradients. From a theoretical perspective, however, the disc effective radius (or equivalently the scale length of the exponential disc) may represent the preferred length scale for normalising the metallicity gradients. The distinction is important because our sample contains bulge + disc systems, and the presence of a bulge affects the measured Petrosian (or S\'ersic) effective radius by making it smaller than the disc effective radius. This effect is particularly important at high masses, where the bulge represents a more significant mass component in galaxies. Observationally, however, the determination of disc effective radius is less robust than the Petrosian radius determination for the MaNGA sample as bulge-disc decompositions are often not unique, and bulges of $\sim$ kpc size may be marginally resolved at the MaNGA median redshift. 

In order to test the effect of using a disc-only effective radius, we cross-matched the MaNGA sample with the \cite{Simard2011} catalogue of bulge-disc decompositions based on the SDSS photometry.\footnote{We use the version of the catalogue where the bulge S\'ersic index is fixed at n=4.} We obtain a positional match for 516 out of the 550 galaxies in our sample. From the Simard catalogue we extract the r-band disc-only effective radii for our sample. 

\begin{figure*} 

\includegraphics[width=0.49\textwidth, trim=0 0 0 0, clip]{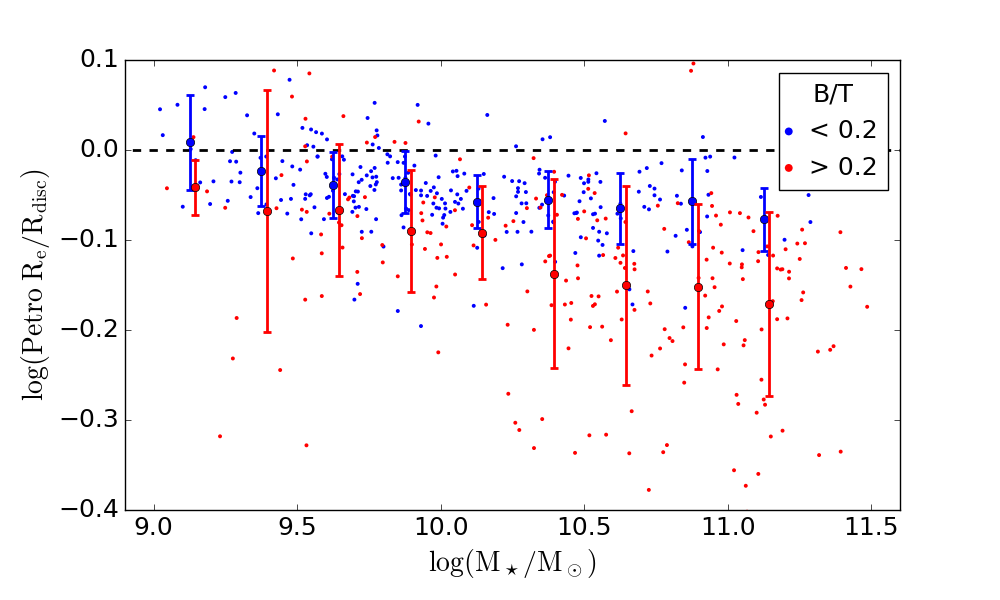}
\includegraphics[width=0.49\textwidth, trim=0 0 0 0, clip]{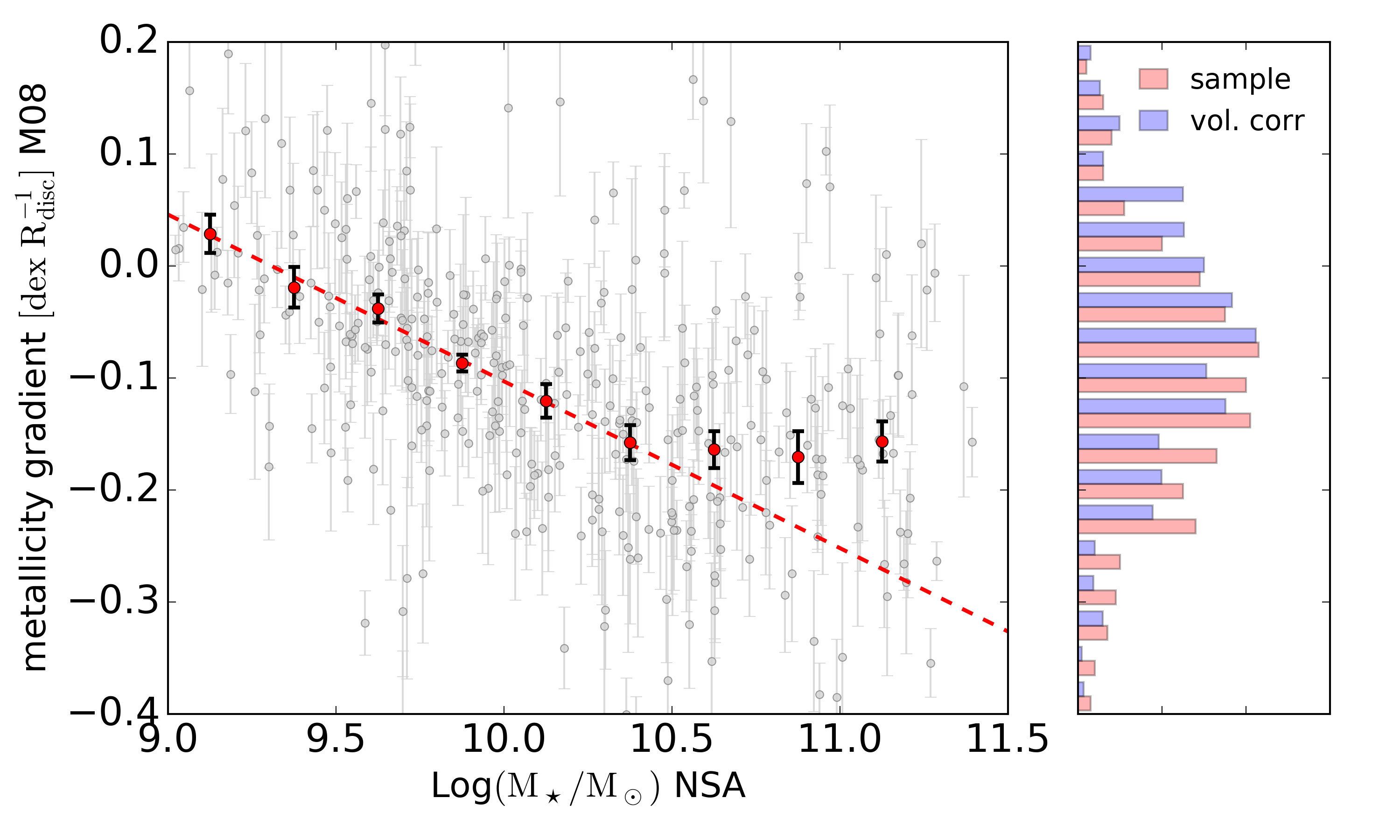}

\caption{\textit{Left}: The difference between the elliptical Petrosian and the disc effective radius as a function of stellar mass. Blue and red points correspond respectively to galaxy with bulge/total (B/T) ratios of $<$ 0.2 and $>$ 0.2. The larger dots with error bars correspond to the median and standard deviations of the two samples as a function of stellar mass. Note that the difference between the two measures of the effective radius increases as a function of B/T.
\textit{Left}: The metallicity gradient (measured in the radial range 0.5 - 2.0 $\rm R_{disc}$ using the M08 calibration based on R23) as a function of stellar mass, using the disc-only effective radius from the \protect\cite{Simard2011} catalogue. The red points represent the median measurements in 0.25 dex stellar mass bins with corresponding errors. The red dashed line is a straight line fit to the median gradient as a function of mass for $\rm log(M_\star/M_\odot) < 10.5$. The histogram represents the distribution for the MaNGA sample (red) and for a volume-weighted sample (blue).}
\label{fig4A.1}

\end{figure*}

In Fig. \ref{fig4A.1} (left) we compare the NSA Petrosian and \cite{Simard2011} disc-only effective radii, colour-coding galaxies according to the bulge/total (B/T) ratio (blue for B/T $<$ 0.2 and red for B/T $>$ 0.5). We note that the disc effective radii are always larger than the elliptical Petrosian effective radii, with the difference increasing as a function of stellar mass at fixed B/T.
Moreover, higher B/T galaxies have a larger difference between the disc-only and elliptical Petrosian effective radii.

In Fig. \ref{fig4A.1} (right) we show the dependence of the metallicity gradient on stellar mass obtained if the disc-only effective radius from \cite{Simard2011} is used instead of the galaxy-wide Petrosian effective radius as a normalisation for the distance. We confirm the qualitative trends observed in Sec. \ref{sec4.1}: a log-linear decrease of the gradient slope with stellar mass, and a flattening of this relation for $\rm log(M_\star/M_\odot) > 10.5$. However, in contrast with the results from Sec. \ref{sec4.1}, the slope of the metallicity gradient does not flatten again for the highest-mass galaxies but remains roughly constant at $\rm \sim -0.16~dex~R_{disc}^{-1}$. Moreover, for all masses the derived gradients are steeper than those obtained normalising by the Petrosian radius, as expected since the Petrosian radius is systematically smaller than the disc effective radius.

For $\rm log(M_\star/M_\odot) < 10.5$ the median metallicity gradient as a function of mass is well-fitted by a straight line relation between the metallicity gradient ($\rm \Delta \log(O/H)$) and stellar mass for the M08 calibration, of the type $\rm \Delta \log(O/H) = a + b~ \log(M_\star/M_\odot)$. We obtain best fit parameters $\rm a=0.15~ (\pm 0.02)$ and $\rm b= - 0.14~ (\pm 0.02)$.

To facilitate comparison with previous work making use of no radial normalisation, we present in Appendix \ref{app:C} the results obtained by calculating the metallicity gradient in units of dex $\rm kpc^{-1}$.

\subsection{The effect of spatial resolution}
\label{sec4A.2}

To further the discussion in Sec. \ref{sec3.3}, in this section we take an empirical approach to quantifying the effect of spatial resolution on the measurement of the metallicity gradient in the MaNGA sample by considering sub-samples differently affected by beam-smearing.

\begin{figure*} 
\includegraphics[width=0.49\textwidth, trim=0 0 0 0, clip]{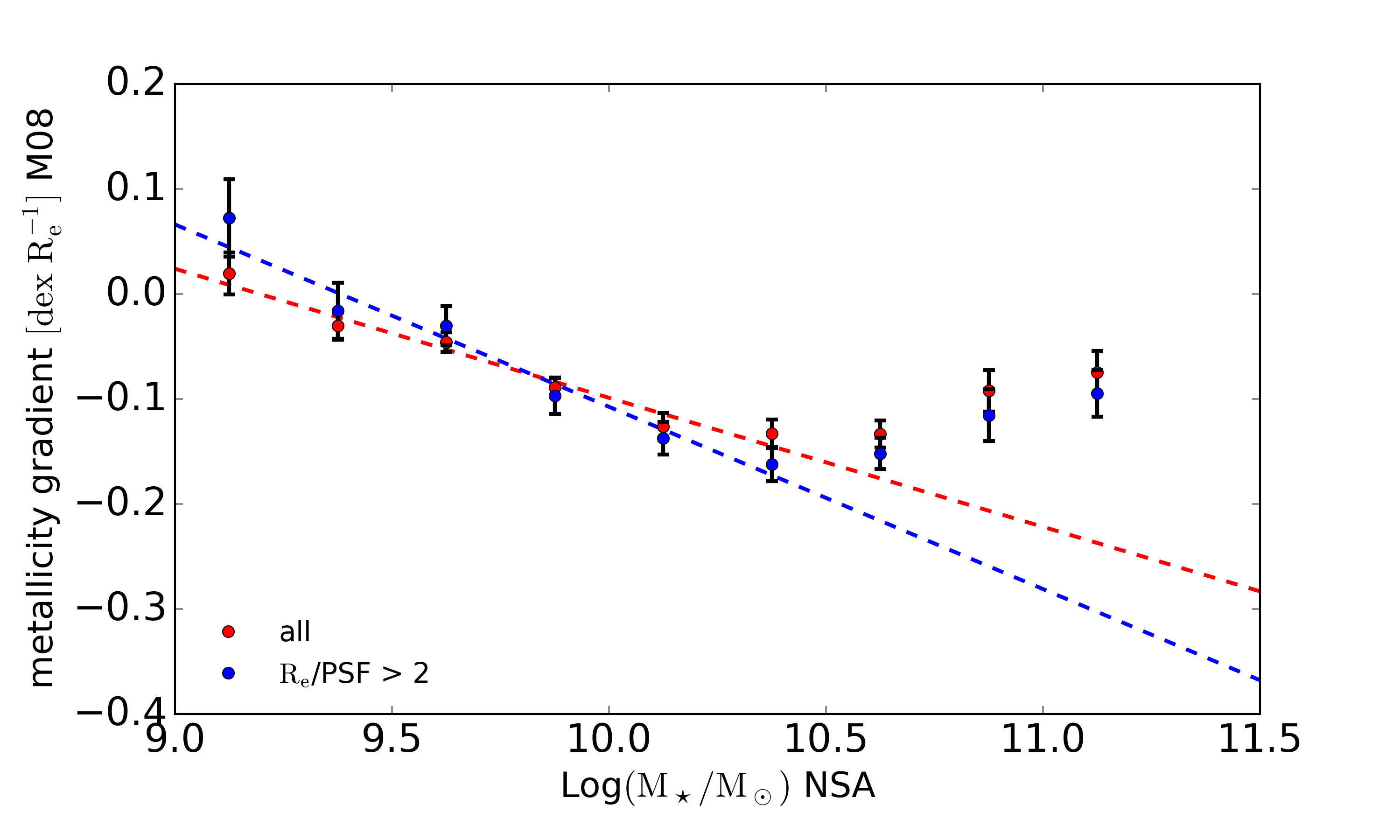}
\includegraphics[width=0.49\textwidth, trim=0 0 0 0, clip]{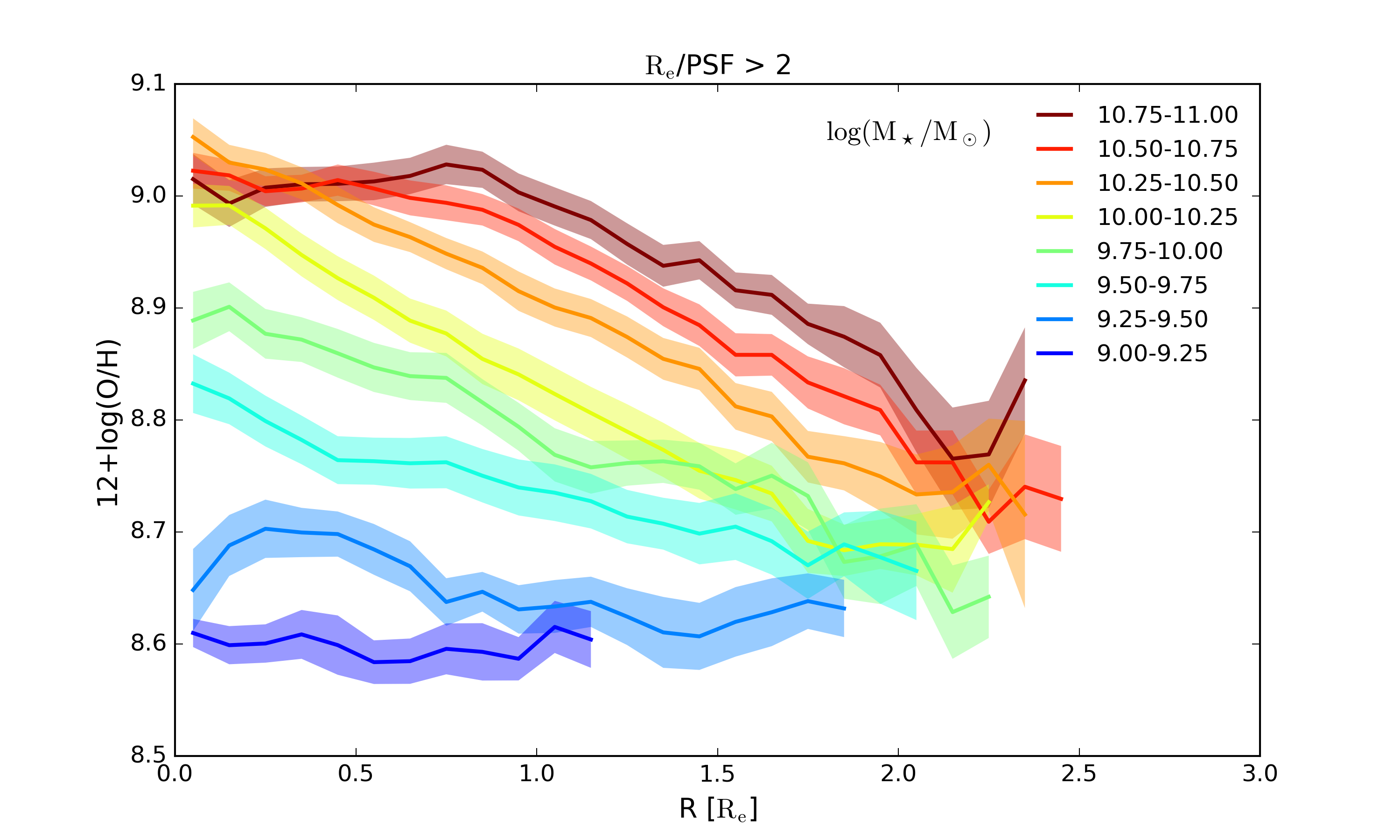}
\caption{ 
\textit{Left}: The dependence of the metallicity gradient on stellar mass using the whole MaNGA sample selected in this work (red) and the the sub-sample of galaxies with $\rm R_e$/PSF $>$2 (blue). Each data point represents the median gradient for the relevant mass bin and the error bars represent the error on the median. The dashed lines represent the best fits linear for the relation in the range $\rm 9.0 < log(M_\star/M_\odot) < 10.5$
\textit{Right}: Stacked metallicity gradients as a function of stellar mass using the sub-sample of galaxies with $\rm R_e$/PSF $>$2.
 }
\label{fig4A.2.1}
\end{figure*}

First, we consider measuring the metallicity gradients for a sub-sample of galaxies with $\rm R_e$/PSF $>$ 2.0. This cut selects 61 \% of the sample (338 galaxies), but biases the selection as it preferentially excludes galaxies observed with the two smallest fibre bundles (19 and 37-fibre IFUs). Selecting MaNGA galaxies based on IFU size generates a surface brightness bias (Wake et al., in prep), since at fixed luminosity MaNGA only observes a narrow redshift range (see Fig. \ref{fig2.1}). Within this redshift range, galaxies observed with larger bundles have lower surface brightness, since they have larger effective radii on the sky. Aware of this potential bias, we perform the same analysis presented in Sec. \ref{sec4.1} and measure the metallicity gradient using the M08 calibration on this sub-sample of well-resolved galaxies. The resulting slopes of the metallicity gradient and stacked metallicity profiles as a function of stellar mass are presented in Fig. \ref{fig4A.2.1}. We observe the sub sample of galaxies with $\rm R_e$/PSF $>$ 2.0 has steeper gradients than the whole sample, but the effect is small and within the statistical error for all mass bins. We note that galaxies with $\rm R_e$/PSF $>$ 2.0 in the lowest mass bin show a marginally inverted gradient. In the right panel of Fig. \ref{fig4A.2.1} we show the stacked metallicity profiles for the $\rm R_e$/PSF $>$ 2.0 and demonstrate that this sub-sample reproduces the main features observed for the whole sample. Namely we further confirm that the flattening of the metallicity gradient in the central regions of massive galaxies is not due to beam smearing effects.

\begin{figure*} 
\includegraphics[width=0.49\textwidth, trim=0 0 0 0, clip]{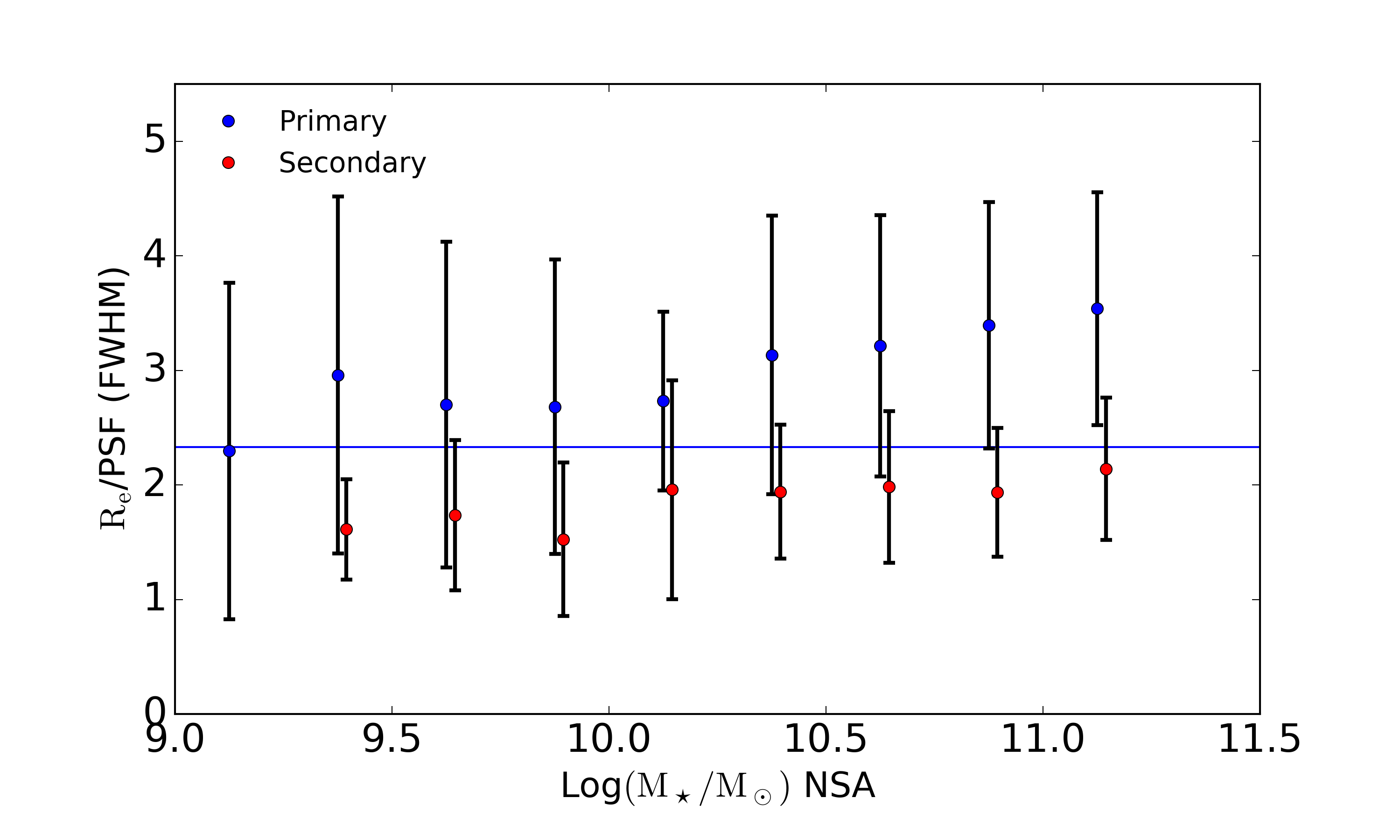}
\includegraphics[width=0.49\textwidth, trim=0 0 0 0, clip]{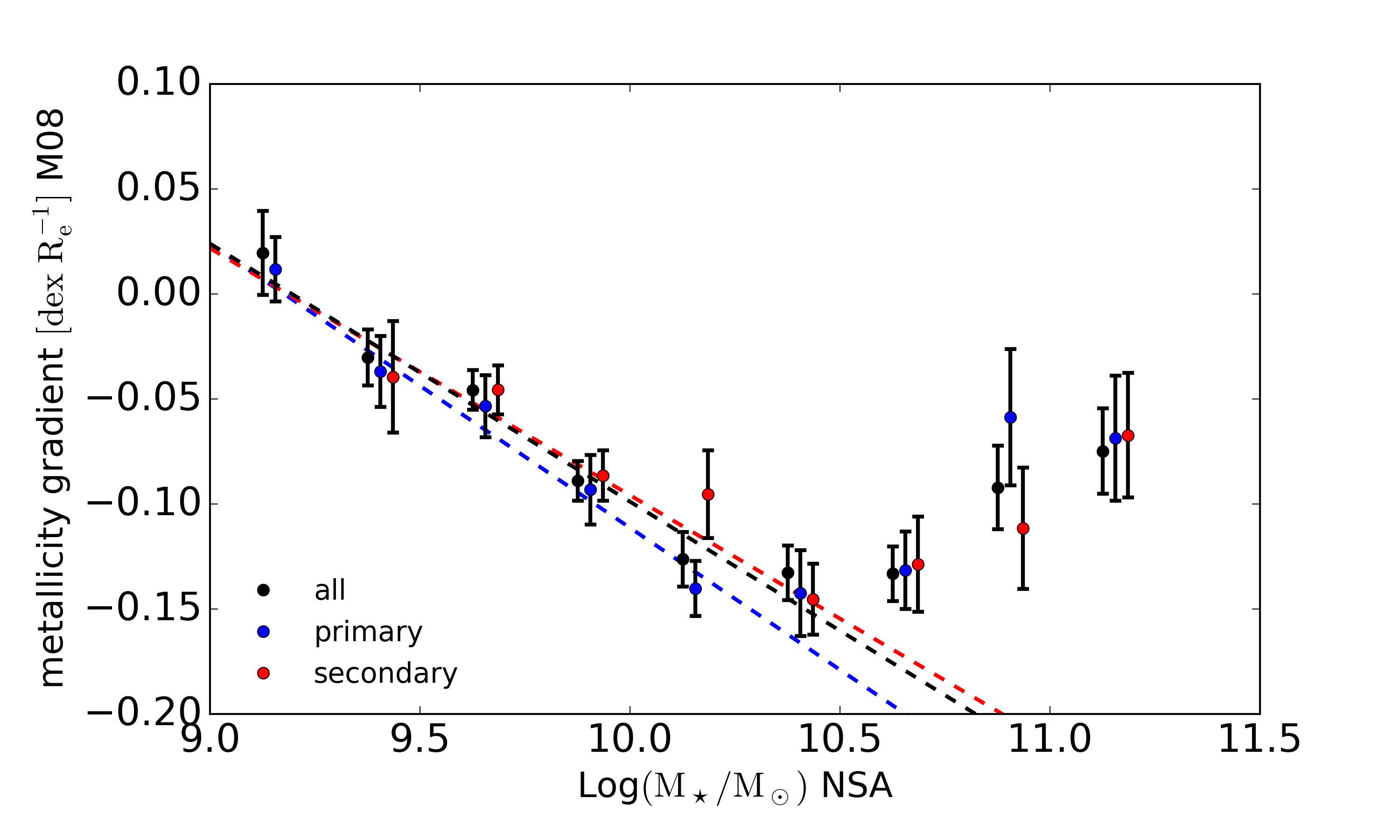}
\caption{ 
\textit{Left}: The dependence of resolution (in terms of $\rm R_e$/PSF) on stellar mass for the primary (blue) and secondary (red) MaNGA samples. At all masses the secondary sample, which covers galaxies out to 2.5 $\rm R_e$, offers lower resolution than the primary sample (covering galaxies out to 1.5 $\rm R_e$).
\textit{Right}: The dependence of the metallicity gradient on stellar mass using the whole MaNGA sample selected in this work (black) and for the primary (blue) and secondary (red) samples separately. Each data point represents the median gradient for the relevant mass bin and the error bars represent the error on the median. The dashed lines represent the best fits for the range $\rm 9.0 < log(M_\star/M_\odot) < 10.5$.
 }
\label{fig4A.2.2}
\end{figure*}

We further test the effect of beam smearing on our results by comparing the metallicity gradients obtained using the primary and secondary samples separately. These two MaNGA sub-samples are selected to both be representative of the local galaxy population, with the secondary sample observing galaxies to larger galactocentric distance and thus affording fewer resolution elements per $\rm R_e$. However for both samples the MaNGA selection uses all bundles sizes at each luminosity (or stellar mass), thus affording a roughly constant number of resolution elements per effective radius as a function of stellar mass. This fact is demonstrated by the left panel of Fig. \ref{fig4A.2.2}, which shows the median $\rm R_e/PSF$ as a function of stellar mass for the primary and secondary samples respectively. 
The resulting median slopes of the metallicity gradient as a function of stellar mass for each sub-sample and for the full MaNGA sample are plotted in the right panel of Fig. \ref{fig4A.2.2}, demonstrating that the secondary sample leads to slopes that are statistically consistent with the primary sample for almost every mass bin. 

In conclusion, the observed trend between the shape of the metallicity gradient and stellar mass appears robust to the effect of resolution.

\section{Comparisons to previous work}
\label{sec5}

\subsection{The metallicity gradient in nearby galaxies}
\label{sec5.1}

\subsubsection{The slope of the metallicity gradient in other IFS surveys}
Recent IFS observations of nearby galaxies \citep{Rosales-Ortega2011, Sanchez2012, Sanchez2014, Ho2015, Menguiano2016, Perez-Montero2016} have provided large datasets suitable for studying the shape of the metallicity gradients in the nearby Universe. To illustrate the source of possible differences with previous work, we compare our results with \citealt{Menguiano2016} (SM16), one of the latest studies of abundance gradients based on the CALIFA survey \citep{Sanchez2012a}, based on a final sample of 129 galaxies. SM16 obtains an average metallicity gradient of $\rm - 0.11\pm 0.07~dex~R_e^{-1}$ using the PP04 calibration (but a lower value of $\rm - 0.08\pm 0.09~dex~R_e^{-1}$ when considering only H\textsc{ii} regions identified via a clump-finding algorithm). This result is comparable in both average slope and scatter with the one presented in this work ($\rm - 0.08~dex~R_e^{-1}$ for both the R23 and O3N2 calibrations), although in case of the directly comparable method for selection of star forming spaxels SM16 reports a slightly steeper gradient.

In order to draw a fair comparison, it is important to note that CALIFA observes more nearby galaxies with a larger IFS bundle, thus resulting in an average of $\sim 11$ resolution elements per $\rm R_e$ (using the effective radii for the sample used in SM16 and a FWHM of 2.5$''$ for CALIFA, as described in \citealt{Garcia-Benito2015}). The CALIFA data is thus much less affected by beam smearing than the data used in this work (which has an average $\rm PSF/R_e = 2.3$). 
A further important difference between this work and SM16 is that the CALIFA sample is not representative for galaxies of $\rm log(M_\star/M_\odot) < 9.65$, because of the adopted diameter-selection strategy \citep{Walcher2014}. This difference in sample selection between the CALIFA and MaNGA samples may lead to a significant discrepancy when comparing the median gradients if, as shown in Sec. \ref{sec4.1}, the metallicity gradient has a dependence on stellar mass. In particular, the larger number of low-mass galaxies in the MaNGA sample is likely to be a significant contribution to the shallower gradient measured by MaNGA with respect to CALIFA. This effect is further accentuated if we apply a volume-correction to the MaNGA sample.

\subsubsection{The stellar mass dependence of the metallicity gradient}
This work presents a clear detection of a non-linear dependence of the metallicity gradient on stellar mass in the nearby Universe (\ref{sec4.1}). This dependence was not clearly detected in the previous, smaller samples, although hints in this direction were already present in the literature \citep{Vila-Costas1992, Zaritsky1994, Garnett1997}. More recently \cite{Perez-Montero2016} make a marginal detection of the dependence between the O/H and N/O gradients and stellar mass, qualitatively similar to the one observed in this work. Namely they observe a trend towards shallower gradient for both log(O/H) and log(N/O) for low-mass (late type) galaxies. Additionally for log(O/H) (but not log(N/O)) they also observe a trend towards shallower gradients for high mass (early type) star forming galaxies. 

As an aside, we note that the results from \cite{Delgado2015} for the stellar metallicity gradients derived from full spectra fitting of the CALIFA data point towards the existence of a similar mass trend as presented in this work (see their Fig. 12 and 13 for a direct comparison). Namely \cite{Delgado2015} obtain the steepest gradients for $\rm log(M_\star/M_\odot) \sim 10.5$ and a gradual flattening for galaxies of lower and higher mass. \cite{Goddard2017b} reach similar conclusions based on a full spectral fitting analysis of the MaNGA data. Namely, they detect a general steepening of the mass-weighted stellar metallicity gradient with stellar mass.
 
We note that the results presented in this paper apply only to star forming galaxies. Early-type galaxies are found to have different metallicity gradient profiles \citep{Pastorello2015, Delgado2015, Goddard2017b}, which are more strongly influenced by the merger history of the system and the accreted stellar component \citep{Coccato2011, Greene2015}.

\subsubsection{The shape of the metallicity gradient at small and large radii}
We confirm, with larger statistics that in previous work, the change in \textit{shape} of the metallicity gradient as a function of radius and mass, namely a flattening at small radii for higher mass galaxies and a mild flattening at large radii especially evident for $\rm log(M_\star/M_\odot) > 10.0$. These features were already detected in smaller samples of galaxies by previous work \citep{Belley1992, Bresolin2009, Werk2010, Werk2011, Rosales-Ortega2011, Bresolin2012} and confirmed by the extensive studies presented by the CALIFA collaboration \citep{Rosales-Ortega2012, Sanchez2014, Menguiano2016}. In addition to confirming the change in shape of the metallicity gradient, in this work we also present evidence for the flattening of the N/O ratio gradient at large radii and the steepening of the N/O in the central regions of massive galaxies. 

It is interesting to note that studies using both CALIFA and MaNGA data have highlighted a correlation between metallicity and local stellar mass surface density \citep{Rosales-Ortega2012, Sanchez2014, Bellesteros2016} and have interpreted galaxy metallicity gradients as a direct result from such a correlation. However, our results clearly show that the stellar surface density cannot be the only driver of metallicity, since the metallicity profiles we observe deviate from exponentials in both the inner regions of massive galaxies and the outer regions of galaxies of all masses. In the case of the inner regions of massive galaxies, the presence of the bulge complicates a direct comparison between metallicity and the disc mass profile. The flattening of the metallicity gradient at large radii, however, may imply a deviation from the relation between mass surface density and metallicity observed across the inner regions of the disc. As discussed further in Sec. \ref{sec6}, the fact that the flattening of the metallicity gradient at large radii is not consistent with a `local' relation between metallicity and stellar mass surface density may be explained invoking feedback processes that lead to the redistribution of metals from the inner to the outer regions.

\subsection{Abundance gradients in the Milky Way}
\label{sec5.2}

A direct comparison of our result with the oxygen abundance gradients measured in the Milky Way is a difficult task for a variety of reasons, including (1) the use of different tracers and calibrations, often tracing different elements, (2) the presence of substantial dust obscuration in the Milky Way disc, (3) the uncertainty on the galactocentric distance of the measured tracers and on the Milky Way scale length itself. 

With regards to metallicity tracers, studies of the present epoch abundance gradient in the Milky Way have been carried out using H\textsc{ii} regions (via measurements of optical emission lines, radio recombination lines and collisionally excited lines, leading to somewhat discrepant results, see for example \citealt{Henry1999, Rudolph2006, Blanc2015}), O/B-type stars \citep{Smartt1997, Rolleston2000}, Cepheids \citep{Luck2003, Pedicelli2009, Lemasle2013, Genovali2014}, open clusters \citep{Magrini2009, Yong2012} and planetary nebulae \citep{Maciel1999, Stanghellini2006}. The results of these studies lead to metallicity gradients generally in the range -0.03 to -0.08 $\rm dex~kpc^{-1}$. 

Given their young ages ($<$ 200 Myr) and their accurate distance measurements, Cepheids are generally considered the best tracers of the present day chemical make-up of the Galaxy. Using the largest available compilation of Cepheid abundance data \cite{Genovali2014} obtain an abundance gradient of -0.06 $\rm dex~ kpc^{-1}$, or -0.18 $\rm dex~R_e^{-1}$ using the Milky Way scale length from  \cite{Bovy2013}.\footnote{\cite{Genovali2014} measure [Fe/H], so the quoted gradient is also affected by the radial dependence of the [O/Fe] ratio.} We can compare this measurement with the gradient predicted from a galaxy of the mass of the Milky Way (assumed here to be $\rm log(M_\star/M_\odot) = 10.8$, \citealt{Bovy2013}) based on the results of this work using the disc-only effective radius (Sec. \ref{sec4A.1}) which is -0.17 $\rm dex~R_e^{-1}$. Considering the systematics involved and the large scatter in the metallicity gradient observed in our sample at each stellar mass, we consider these values to be in excellent agreement.

Studies of the abundance gradient in our Galaxy based on different tracers points towards the possibility of a flattening of the metallicity gradient at large radii ($\rm R> 10~kpc$, see for example \citealt{Pedicelli2009} using Cepheids, \citealt{Vilchez1996} using H\textsc{ii} regions or \citealt{Magrini2009} using open clusters), although this claim is contested by other authors using the same tracers (see, for example, the discussion in \citealt{Stasinska2012}). Using the scale length from \cite{Bovy2013}, the observed flattening in the Milky Way metallicity gradient would occur at $\rm \sim 2.8~R_e$, a larger galactocentric distance than the flattening detected in this work for $\rm R>2.0~R_e$. Unfortunately MaNGA data does not cover the radial range $\rm R > 3~R_e$, so we are currently unable to demonstrate whether the mild flattening detected in this work carries on at even larger radii. Interestingly, in the Milky Way a flattening is also reported for the N/O ratio from studies of H\textsc{ii} regions \citep{Vilchez1996}.



\section{Discussion}
\label{sec6}


\subsection{Chemical evolution models and `inside-out' growth}

Chemical evolution studies have provided some of the earliest evidence in support of the theoretical framework of disc formation in the cosmological framework and `inside-out' growth \citep{Chiosi1980, Matteucci1989, Boissier1999, Chiappini2001}, since the difference in assembly time across the galactic disc naturally produces a metallicity gradient. The inside-out growth scenario is further supported by observations of colour gradients \citep{Bell2000, Munoz-Mateos2007} and spectroscopically-inferred age profiles of disc galaxies \citep{Perez2013, Delgado2015, Ibarra-Medel2016}. 

It is important to note that other factors beyond the radial variation in assembly time can play a crucial role in determining the metallicity gradient. As demonstrated by \cite{Goetz1992} from a purely analytical perspective, metallicity gradients can also be generated by a radial dependence of the metal yield (which, assuming that nucleosynthetic physics is universally constant, implies a change in the stellar initial mass function, IMF) or a radial dependence of the star formation efficiency (SFE= SFR/M$\rm _{gas}$). Leaving aside any possible IMF variation, different assumptions regarding the radial variation of the star formation efficiency have been investigated by classical chemical evolution models. These differences in the parametrisation of the SFE are not merely a technical detail: they are at the core of the different predictions for the time evolution of the metallicity gradient presented in the literature. In particular, models where the SFE is assumed constant as a function of radius generally predict a steepening of the metallicity gradient with time \citep{Matteucci1989, Chiappini2001}, while models that assume a decrease of the SFE with radius generally predict a flattening with time (\citealt{Boissier1999, Molla2005}, although this distinction is not always so clear cut, see for example the discussion in \citealt{Mott2013}).

On top of these effects, radial flows and gas recycling from the hot halo (which is dependant on the overall efficiency of ejective feedback) can act to flatten (or, albeit for rather contrived choices of parameter, steepen) pre-existing metallicity gradients \citep{Lacey1985, Schonrich2009, Pezzulli2016}. Recent progress in semi-analytical modelling \citep{Fu2009} and hydrodynamical simulations performed in a cosmological context \citep{Pilkington2012, Gibson2013} have demonstrated the fundamental role that metal recycling via galactic fountains has on the metallicity gradient. These effects of feedback and recycling ought to be more extreme in low-mass galaxies, where outflows play a larger role in regulating the star formation and gas supply.

It is beyond the scope of this work to revisit the predictions of classical chemical evolution models in light of the new observations presented here. We wish, however, to comment on the fact that the clear detection of a steepening of the metallicity gradient with stellar mass (between 0.5 $\rm R_e$ and 2.0 $\rm R_e$) has important consequences for our understanding of the time evolution of metallicity gradients. In a simple model of chemical evolution, chemical enrichment, time and mass growth are tightly linked. In this sense lower mass galaxies can be interpreted as the `progenitors' of the higher mass star forming galaxies (note that this equivalence is only true in a limited sense, since the physical conditions of low-mass high-redshift galaxies are assumed to be different from the low-redshift equivalents). In a simple inside-out growth model (with no radial flows and no wind recycling) and a SFE profile which decreases as a function of radius (as observed in local galaxies, \citealt{Leroy2008}) the metallicity gradient should initially (i.e. in low-mass galaxies) be steep, as star formation occurs primarily in the central regions, and subsequently flatten, as star formation proceeds in the outer regions of the disc. Moreover, in this scenario, the flattening should occur `inside-out', with central regions reaching their equilibrium metallicity earlier than outer regions (see Sec. \ref{sec:chem_eq} and next subsection), as is indeed observed in this work. Hence, some contribution from ejective feedback may be necessary to reproduce the flat abundance gradients observed in the lowest-mass galaxies in our sample. In these galaxies the shallower gravitational potential well may favour the launching of large scale galactic winds, reducing the amount of low angular momentum gas that turns into stars in the central regions of the disc \citep{Brook2012}.

\subsection{Interpreting the shape of the metallicity gradient}
\label{sec:chem_eq}

It is a general property of chemical evolution models accounting for infall, star formation and outflows to evolve towards an equilibrium metallicity at late times. The concept of an equilibrium chemical abundance has been discussed extensively in recent literature focussing on the study of `bath-tub' chemical evolution models \citep{Dave2012, Lilly2013, Peng2015, Belfiore2016}. In these models, the flattening of the mass-metallicity relation at high masses \citep{Tremonti2004} and the flattening of the metallicity gradient in the central regions of massive galaxies can be ascribed to the general behaviour of evolved systems to reach an equilibrium abundance at late times, corresponding to low gas fractions. Given the time delay between the production of oxygen and nitrogen, the nitrogen abundance keeps increasing even if the global metallicity reaches an equilibrium value, resulting in the different behaviours of the oxygen and N/O gradients in the central regions of massive galaxies. Interestingly, gradual dilution of the interstellar medium in the central regions of massive galaxies is predicted by some models \citep{Yates2014} and might contribute to explaining the observed flattening and/or marginal inversion of the metallicity gradient in the very central regions of the most massive galaxies in our sample, while not affecting our conclusion for the N/O ratio.

The flattening of the metallicity gradients at large galactocentric radii deserves a more detailed discussion. The outer regions of galaxies are an ideal laboratory to study disc assembly at the current epoch, since these regions are characterised by extreme conditions: high gas fractions, long dynamical timescales and low stellar surface density. These properties indicate that outer disc regions are relatively un-evolved, a realisation that stands in contrast with the relatively high observed chemical abundances (8.7 <12+log(O/H) <8.9, for the M08 R23 calibration). Indeed, our data shows that at large radii the profiles of galaxies of different masses tend to converge to a common metallicity, similar in value to the metallicity observed throughout the disc of the low-mass galaxies in our sample. This value of metallicity is much higher than than the metallicity floor of $\sim$ 1\% $\rm Z_\odot$, found in metal-poor galaxies in the local Universe \citep{Izotov2006, Morales-Luis2011}, which is expected to be set by direct accretion of gas from the cosmic web at z$\sim$0 \citep{Danforth1994, Arnaud1994, Almeida2014}.

The common metallicity observed in the MaNGA data at large radii may reflect the fact that chemical enrichment initially happens quickly, especially for systems with a high star formation efficiency (see for example Fig. 5 in \citealt{Vincenzo2016}). Outer discs, however, are likely characterised by low star formation efficiencies and previous work has argued against the self-enrichment scenario, pointing out that the current level of star formation is insufficient to explain the observed high abundances, even when extrapolated back for a Hubble time \citep{Bresolin2012}.

As an alternative to in-situ production, metals in the outer regions of the disc may originate from the highly-enriched regions of the inner disc through either radial flows or re-accretion of enriched coronal gas. In the enriched infall scenario, the observed metals at large radii are most likely the result of a `wind recycling' process, where the halo gas is progressively enriched by outflows from the more star forming inner regions of the disc, cools and falls back onto the disc over a $\sim$ 1 Gyr timescale, as predicted in recent hydrodynamical simulations \citep{Oppenheimer2010, Christensen2016, Muratov2016, Armillotta2016}. This scenario, however, does not provide a natural interpretation of the common metallicity observed in MaNGA galaxies at large radii.  Further modelling of galactic fountains and their effect of chemical abundances may help to shed more light on this question.

\subsection{The N/O ratio gradient}
\label{sec6.3}

The nucleosynthetic origin of nitrogen is more complex than that of oxygen. At low metallicity, nitrogen is assumed to be produced with a yield that does not depend on metallicity (so called `primary' nucleosynthetic origin). At higher metallicity, however, nitrogen is synthesised by low and intermediate mass stars, with a yield proportional to the initial metallicity of gas from which these stars form (`secondary' nucleosynthetic origin). The metallicity dependence of the nitrogen yield, combined with the time delay due to the lifetime of its stellar progenitors, are responsible for the non-linear relation between O/H and N/O, observed in nearby H\textsc{ii} regions \citep{Alloin1979, Vila-Costas1993, Henry2000}, integrated spectroscopy of local and high redshift galaxies \citep{Perez-Montero2012, Vincenzo2016, Strom2016, Masters2016} and IFS spectroscopy of local galaxies \citep{Perez-Montero2016}, including in this work.

Chemical evolution models taking into account stellar lifetimes and metallicity-dependant yields predict that the nitrogen abundance radial gradient ought to be steeper than that for O/H. The prediction is a direct consequence of the time delay between oxygen and nitrogen production, causing the difference in the production rate of nitrogen between the inner and the outer regions of the disc to be even larger than that for oxygen \citep{Matteucci1989}. Another important prediction of modelling the N/O ratio is that the position of systems in the O/H versus N/O plane depends on the ratio of current to past star formation, or, in the context of a self-consistent chemical evolution model, on the SFE \citep{Molla2006, Vincenzo2016}. Lower SFE delays the production of nitrogen at fixed oxygen abundance, thus causing low SFE regions to lie at higher N/O at fixed O/H. 

The results from Sec. \ref{sec4.6} can thus be interpreted as a consequence of the different star formation efficiency between the central and the outer regions of discs. For example, in Fig. \ref{fig4.6} we observe that the central regions of galaxies in the mass range $\rm log(M_\star/M_\odot) = 9.5-9.75)$ have the same oxygen abundance as the outer disc ($\rm R=1.75-2.0~R_e$) of galaxies in the mass range $\rm log(M_\star/M_\odot) = 10.5-10.75)$. However the N/O ratio corresponding to the outer disc of the higher mass bin is $\sim$ 0.15 dex higher. This effect can be ascribed to a systematically lower SFE in the outer discs.

Metal redistribution via wind recycling, which we have already considered in the context of the flattening of the O/H gradient at large radii, may also have an effect in the N/O versus O/H plane. Indeed, higher N/O ratios in the outer disc of massive galaxies can also be interpreted as a consequence of the pollution of galaxy outskirts with some amount of high-metallicity, high-N/O gas from the central regions. Upon mixing with the lower-metallicity gas at larger radii, the metallicity of the infalling gas is significantly diluted, while its N/O ratio only decreases slightly in mixing because of the highly non-linear relation between O/H and N/O (see for example the models presented in \citealt{Koppen2005,Belfiore2015}). In this case the N/O ratio acts as a tracer of the original metallicity of the gas polluting the outer regions. Further modelling, taking into account the combined effect of radial variations in SFE and metal redistribution will be needed in order to fully exploit the information encoded in the O/H versus N/O plane.

\section{Summary and conclusions}
\label{sec7}

We have investigated the gas phase metallicity and nitrogen abundance gradients for a sample of 550 local galaxies, spanning the stellar mass range $\rm 10^9-10^{11.5}~M_{\odot}$, by exploiting integral field spectroscopy from the SDSS-IV MaNGA (data release 13). We select star forming regions based on the [SII] BPT diagram and the \cite{Kewley2001} demarcation line. Metallicity is calculated making use of the strong line calibration of \cite{Maiolino2008}, based on R23, and \cite{Pettini2004}, based on O3N2. The N/O ratio is estimated from the N2O2 ratio following the calibrations proposed by \cite{Pagel1992} and \cite{Thurston1996}. The effect of galaxy inclination and spatial resolution effect have been taken into account, by excluding galaxies highly inclined ($>$60$^\circ$) and by focusing on the metallicity gradients outside the central 0.5$\rm ~R_e$. We summarise the main findings below.

\begin{enumerate}

\item The metallicity gradient (in the radial range 0.5 - 2.0~$\rm R_e$) is flat for low mass galaxies ($\rm M_\star <10^{9.0} ~M_{\odot}$), steepens for more massive galaxies until $\rm M_\star \sim10^{10.5} ~M_{\odot}$ and then flattens slightly again for even more massive systems. This behaviour is confirmed when using the disc-only effective radius as a radial scale length, and is robust to the effects of resolution. The gradient of the N/O abundance ratio shows a very mild steepening as a function of galaxy mass over the whole mass range considered.

\item The most massive galaxies (M$\rm \sim 10^{11}~M_{\odot}$) are characterised by a flattening of the metallicity gradient in the central ($\rm R< 1.0~R_e$) regions, corresponding to a steepening of the N/O ratio gradient in the same radial range.

\item A mild flattening of the metallicity gradient is observed for R$>$2~R$\rm _e$, especially evident in stacked profiles of galaxies with masses $\rm > 10^{10} ~M_\odot$. Such a flattening at large radii is also observed in stacked profiles of the N/O ratio profile.

\item We find a secondary dependance of the metallicity gradient on concentration and sSFR for galaxies of high concentration (C$>$ 2.6) and low sSFR (log(sSFR) $<$ -1.5 $\rm Gyr^{-1}$), in the sense that galaxies of higher concentration and lower sSFR have flatter gradients than expected from their stellar mass.

\item Spatially resolved regions follow the expected relation between N/O and O/H. The scatter in this relation is found to depend on radius. At fixed O/H, outer regions of massive discs have higher nitrogen abundance N/O than the central region of low mass galaxies.

\end{enumerate}

We have discussed the likely interpretation of these trends and their implications for our understanding of disc formation and chemical evolution. Specifically we find that

\begin{itemize}

\item The flattening of the metallicity gradient in massive galaxies, and especially in central regions, can be explained as a consequence of metallicity having reached the equilibrium value, generally attained in evolved, gas poor systems.

\item The flat gradient observed in low mass galaxies supports the need for strong feedback, gas mixing and wind recycling in low-mass systems.

\item Wind recycling and lower SFE can in combination explain the relatively high abundances observed at large radii and the position of regions in the outer disc in the N/O versus O/H diagram. Neither scenario, however, naturally explains why metallicities at large radii tend to a common value (12+log(O/H) $\sim$ 8.6-8.8 using the R23 calibration). Further abundance data at large radii and future hydrodynamical simulations including chemical enrichment may help shed more light on this issue.

\end{itemize}

Overall, this work emphasises that the quality of the chemical abundance data from large IFS surveys like MaNGA allows us to go beyond the description of abundance gradients as simple exponential profiles. The deviations from an exponential profile observed in the inner and outer regions of galaxies provide crucial new constraints for chemical evolution models and their implementation in hydrodynamical simulations, particularly on the relative strength of feedback and wind recycling.


\section*{Acknowledgements}
\begin{small}
F.B. and R.M. acknowledge funding from the United Kingdom Science and Technology Facilities Council (STFC). R.M. acknowledges funding from the European Research Council (ERC), Advanced Grant 695671 `QUENCH'. S.F.S. thanks the CONACYT-125180, DGAPA-IA100815 and DGAPA-IA101217 projects for providing him support in this study. C.A.T. acknowledges support from National Science Foundation of the United States grant 1412287 and 1554877. This research made use of {\tt Marvin}, a core Python package and web framework for MaNGA data, developed by Brian Cherinka, Jos\'e S\'anchez-Gallego and Brett Andrews, 

This work makes use of data from SDSS-IV. Funding for SDSS has been provided by the Alfred P.~Sloan Foundation and Participating Institutions. Additional funding towards SDSS-IV has been provided by the U.S. Department of Energy Office of Science. SDSS-IV acknowledges support and resources from the Centre for High-Performance Computing at the University of Utah. The SDSS web site is {\tt www.sdss.org}.

SDSS-IV is managed by the Astrophysical Research Consortium for the Participating Institutions of the SDSS Collaboration including the  Brazilian Participation Group, the Carnegie Institution for Science, Carnegie Mellon University, the Chilean Participation Group, the French Participation Group, Harvard-Smithsonian Center for Astrophysics, Instituto de Astrof\'isica de Canarias, The Johns Hopkins University, Kavli Institute for the Physics and Mathematics of the Universe (IPMU) / University of Tokyo, Lawrence Berkeley National Laboratory, Leibniz Institut f\"ur Astrophysik Potsdam (AIP),  Max-Planck-Institut f\"ur Astronomie (MPIA Heidelberg), Max-Planck-Institut f\"ur Astrophysik (MPA Garching), Max-Planck-Institut f\"ur Extraterrestrische Physik (MPE), National Astronomical Observatory of China, New Mexico State University, New York University, University of Notre Dame, Observat\'ario Nacional / MCTI, The Ohio State University, Pennsylvania State University, Shanghai Astronomical Observatory, United Kingdom Participation Group, Universidad Nacional Aut\'onoma de M\'exico, University of Arizona, University of Colorado Boulder, University of Oxford, University of Portsmouth, University of Utah, University of Virginia, University of Washington, University of Wisconsin, Vanderbilt University, and Yale University.

\textit{The MaNGA data used in this work is part of SDSS data release 13 \citep{SDSS_DR13}, publicly available at {\tt http://www.sdss.org/dr13/manga/manga-data/}.}
\end{small}


\bibliography{bib16}
\bibliographystyle{mnras}


\appendix

\section{The determination of disc inclination}
\label{app:B}

\begin{figure} 
\includegraphics[width=0.49\textwidth, trim=0 0 0 0, clip]{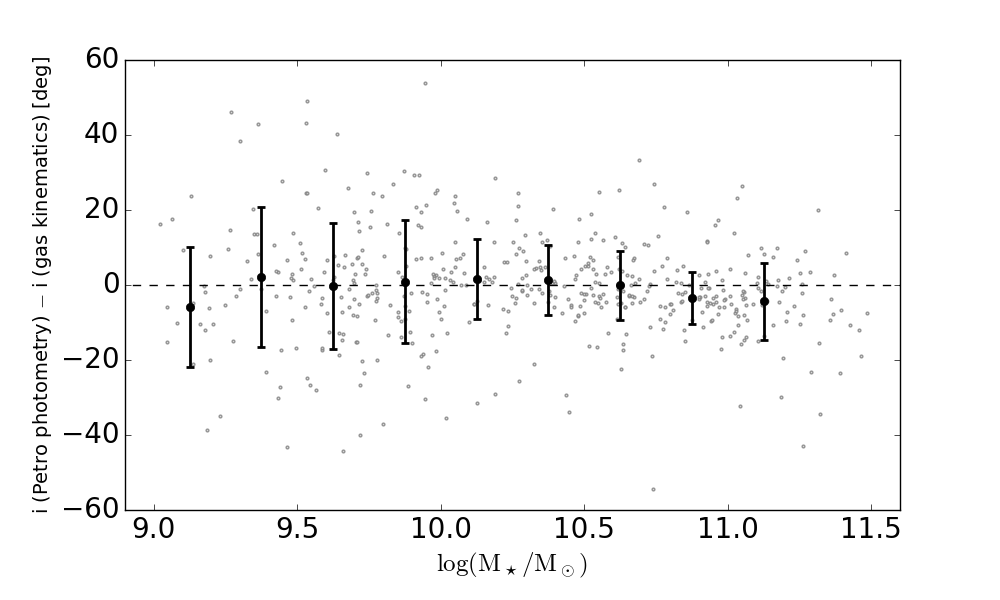}
\caption{The difference between the inclination calculated from elliptical Petrosian photometry and that obtained from fitting the gas kinematics with a thin-disc model (see Westfall et al., in prep). The black points show the median in 0.25 dex mass bins, with the respective standard deviation represented by the error bars.}
\label{figB2}
\end{figure}

In this work the inclination of the disc is derived from the axis ratios (b/a) obtained by performing elliptical Petrosian photometry (from the NSA catalogue). In order to estimate the effect of errors on the inclination on the metallicity gradient, we use the toy model galaxy described in Sec. \ref{sec3.3}. We set up models with different inclinations (0, 30, 45, 60, 75 $^\circ$) and recover the slope of metallicity gradient assuming the measured inclination is affected by a Gaussian error of 10$^\circ$. The resulting difference between the true and measured slope of the metallicity gradient after 1000 noise realisations are computed. Even though the distributions have a positive tail (i.e. a tail where the inferred slope is flatter than the measured one), their median agrees with zero (no systematic difference between measured and true slope) to better than 0.01 $\rm dex~R_e^{-1}$. We therefore conclude that random errors on the inclination consistent with those expected from the photometric determination do not affect the results presented in this paper.

A more serious concern is represented by possible systematic biases in the determination of b/a. For low mass galaxies ($\rm log(M_\star/M_\odot) < 9.5$) the axis ratio can be ill-defined due to clumpiness, lopsidedness and generally weak continuum heavily contaminated by line emission. At the high mass end ($\rm log(M_\star/M_\odot) > 10.5$), on the other hand, the bulge becomes a concern, as light from the central spheroidal component may bias the measurement of the disc b/a. In order to prove that these effects do not bias the inclinations measured from photometry for our sample we have compared the Petrosian inclinations with the inclinations derived by fitting a thin-disc model to the gas velocity field (details in Westfall et al., in prep.). Fig. \ref{figB2} shows the comparison between the kinematic and Petrosian inclinations, demonstrating that the two estimates of inclination are in good agreement over the whole mass range, despite significant dispersion ($\sim 16^\circ$ at the low mass end to $\sim 19^\circ$ at the high mass end). We conclude that the systematic median difference between Petrosian and kinematic inclination is always small (< 6$^\circ$), even in low mass systems.

\section{The metallicity gradient in units of dex/kpc.}
\label{app:C}

\begin{figure} 
\includegraphics[width=0.49\textwidth, trim=0 0 0 0, clip]{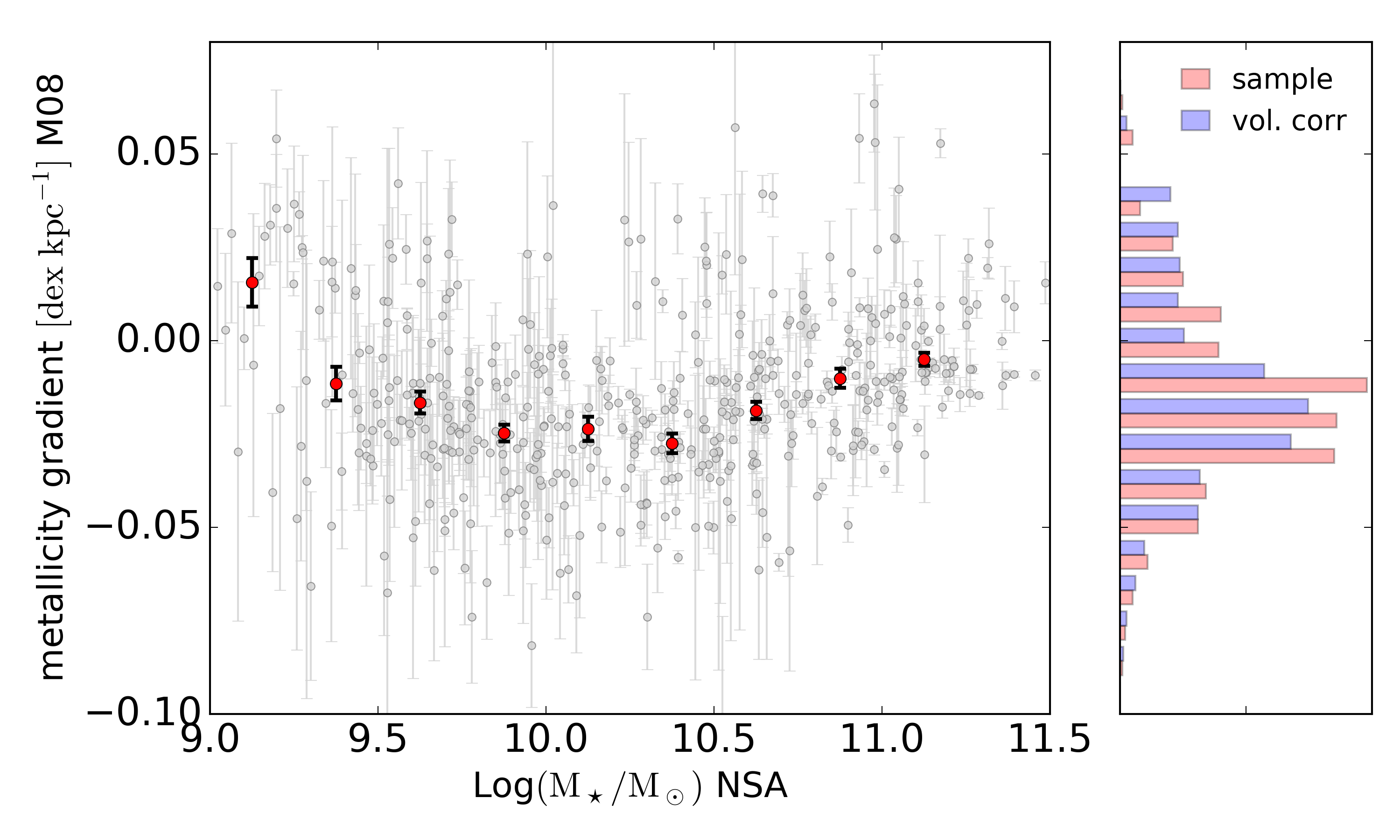}
\includegraphics[width=0.49\textwidth, trim=0 0 0 0, clip]{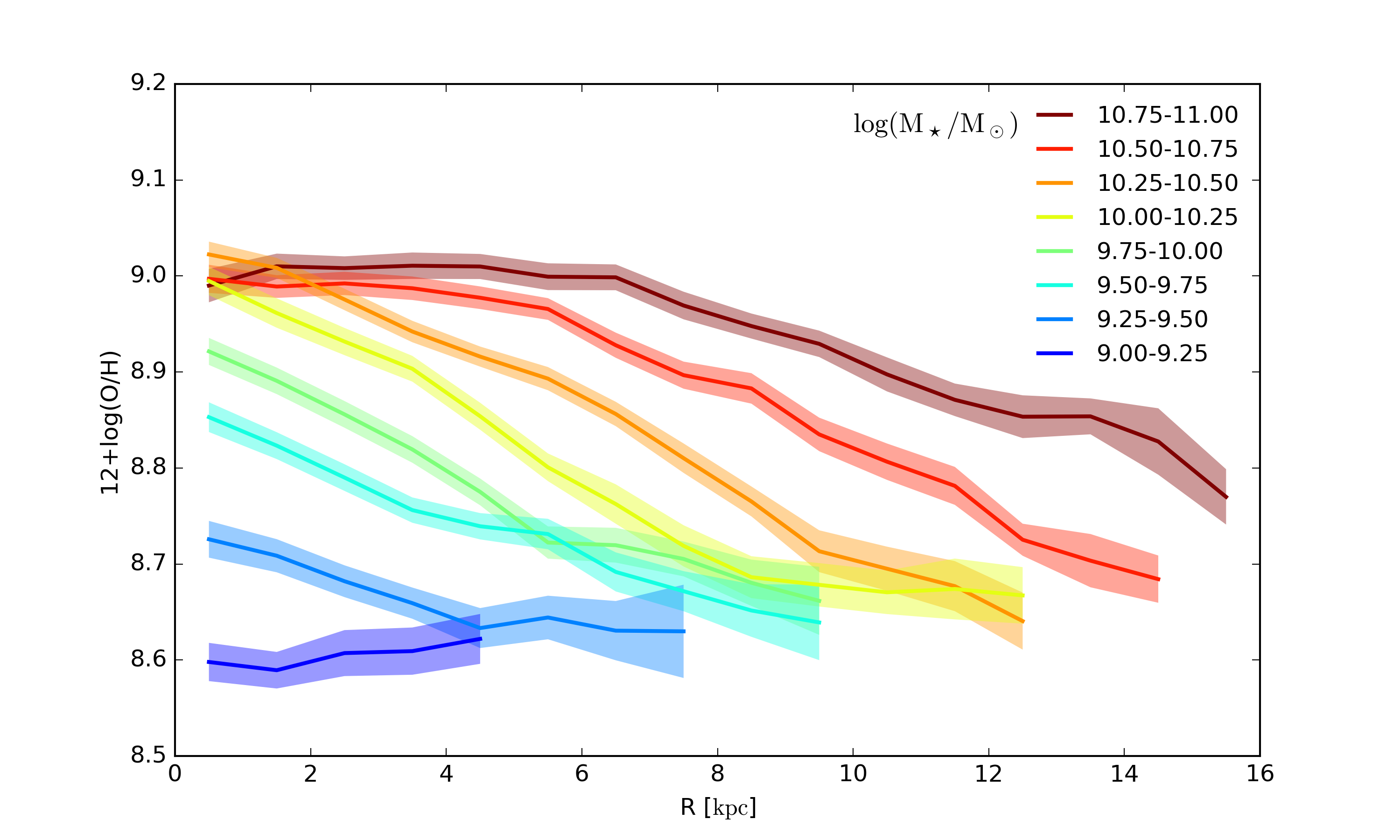}
\caption{Same as Fig. \protect\ref{fig4.1}, but calculating the metallicity gradient in units of dex $\rm kpc^{-1}$.}
\label{figAPPC}
\end{figure}

To facilitate comparisons with previous work on abundance gradient which did not adopt any scale length normalisation we present in this Appendix the results obtained measuring the metallicity gradient in units of dex $\rm kpc^{-1}$. 
We warn the reader, however, that dropping any radial normalisation introduces a \textit{strong} dependence of beam smearing on stellar mass, hence making it more difficult to interpret mass trends in the MaNGA data. Moreover, since in MaNGA galaxies are observed out to a fixed radial distance in units of $\rm R_e$, galaxies of different masses are covered to different physical distances. Hence in this Appendix the metallicity gradient is fitted with a straight line over the whole available radial range, as it is not clear how to self-consistently define an `inner' and `outer' disc in terms of physical distance across the wide stellar mass range we are studying. 

The results, summarised in Fig. \ref{figAPPC}, demonstrate that several features of the analysis presented in this work remain valid when metallicity gradients are measured in units of dex $\rm kpc^{-1}$. Namely, gradients for $\rm log(M_\star/M_\odot) \sim 9.0$ are flat or marginally inverted, the metallicity gradient steepens with stellar mass until $\rm log(M_\star/M_\odot) \sim 10.5$, while becoming flatter again, but still negative, for higher masses. The change in shape of the metallicity gradient as a function of radius and mass is also evident in the metallicity stacks (lower panel of Fig. \ref{figAPPC}), with higher mass galaxies showing a clear flattening in the central regions and intermediate mass galaxies displaying a flattening at large radii.

\bsp
\label{lastpage}
\end{document}